\documentclass[onecolumn,usenatbib]{mnras}

\usepackage{newtxtext,newtxmath}

\usepackage[T1]{fontenc}

\DeclareRobustCommand{\VAN}[3]{#2}
\let\VANthebibliography\thebibliography
\def\thebibliography{\DeclareRobustCommand{\VAN}[3]{##3}\VANthebibliography}

\usepackage{graphicx}	
\usepackage{amsmath}	

\usepackage{multicol}

\usepackage{newtxtext,newtxmath}

\def\M{M$_\odot$}
\def\L{L$_\odot$}

\title[SLSNe in review]{Superluminous supernovae: an explosive decade}

\author[Matt Nicholl]{
Matt Nicholl$^{1}$\thanks{E-mail: m.nicholl.1@bham.ac.uk}
\\
$^{1}$Institute for Gravitational Wave Astronomy and School of Physics and Astronomy, University of Birmingham, Birmingham B15 2TT, UK
}

\date{Invited review for Astronomy \& Geophysics (RAS Journals)}

\pubyear{2021}

\begin{document}
\label{firstpage}
\pagerange{\pageref{firstpage}--\pageref{lastpage}}
\maketitle

\begin{abstract}
I review our current understanding of superluminous supernovae, mysterious events 100 times brighter than conventional stellar explosions.
\end{abstract}





\section{Introduction: the discovery of SLSNe}

The study of supernovae spans centuries and cultures, with Chinese astronomers recording the oldest known `guest star' in 185 AD. By the turn of the millenium, it was well established that core-collapse supernovae (CCSNe) signal the death of a star of more than 8\,\M\ (where \M\ is the solar mass). When the nuclear reactor in its core has converted all its fuel into stable iron, and can no longer extract energy to support itself, it collapses to a neutron star and releases its gravitational energy. Type Ia supernovae (SNe Ia) occur when a white dwarf gains sufficient mass from a binary companion to encounter a runaway thermonuclear reaction. Both types of supernova release $\sim 10^{51}$\,erg in kinetic energy, peak with a luminosity up to $10^9$\,\L, and produce many of the chemical elements needed for life in the Universe.

\begin{figure*}
    \centering
    \includegraphics[width=5.8cm]{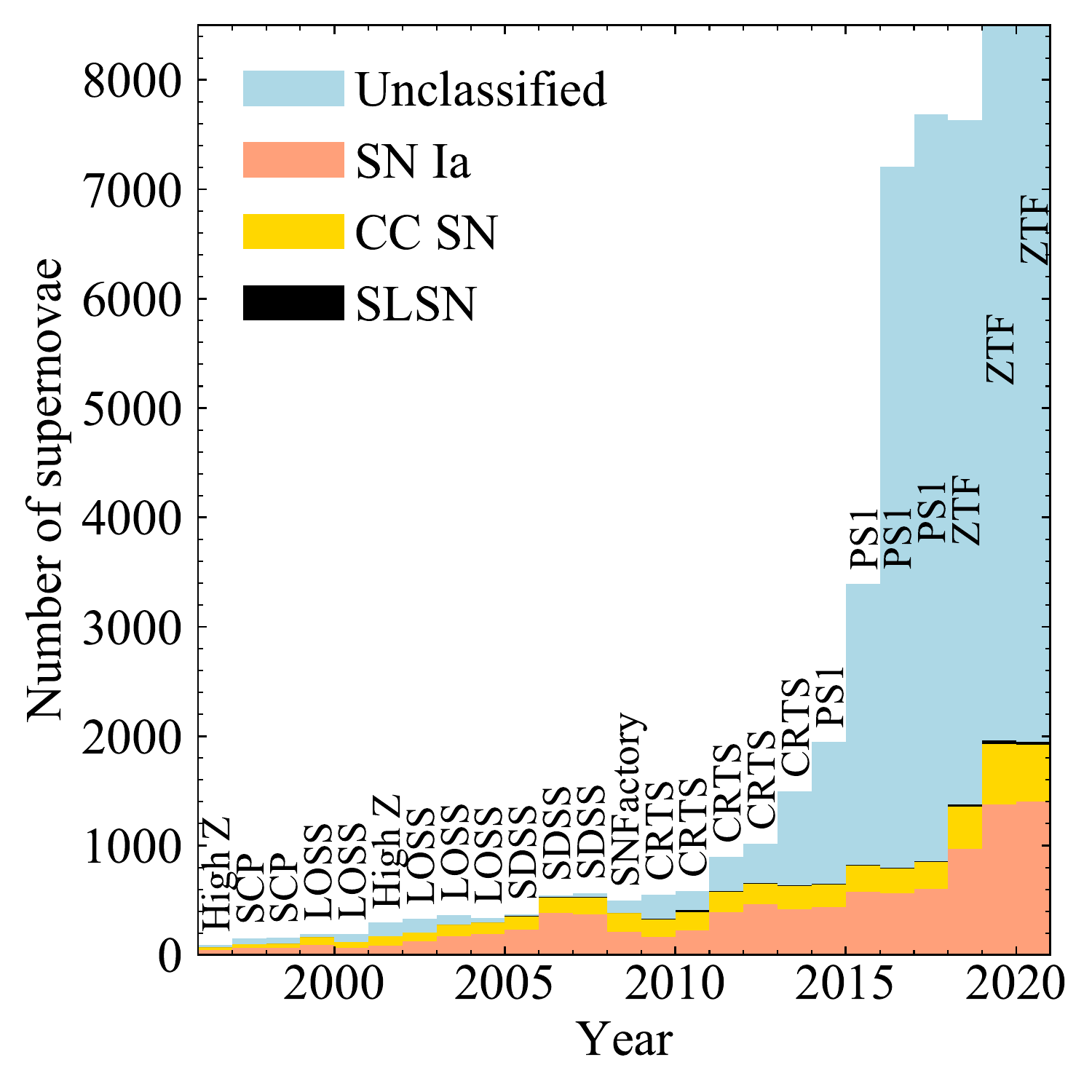}
    \includegraphics[width=5.8cm]{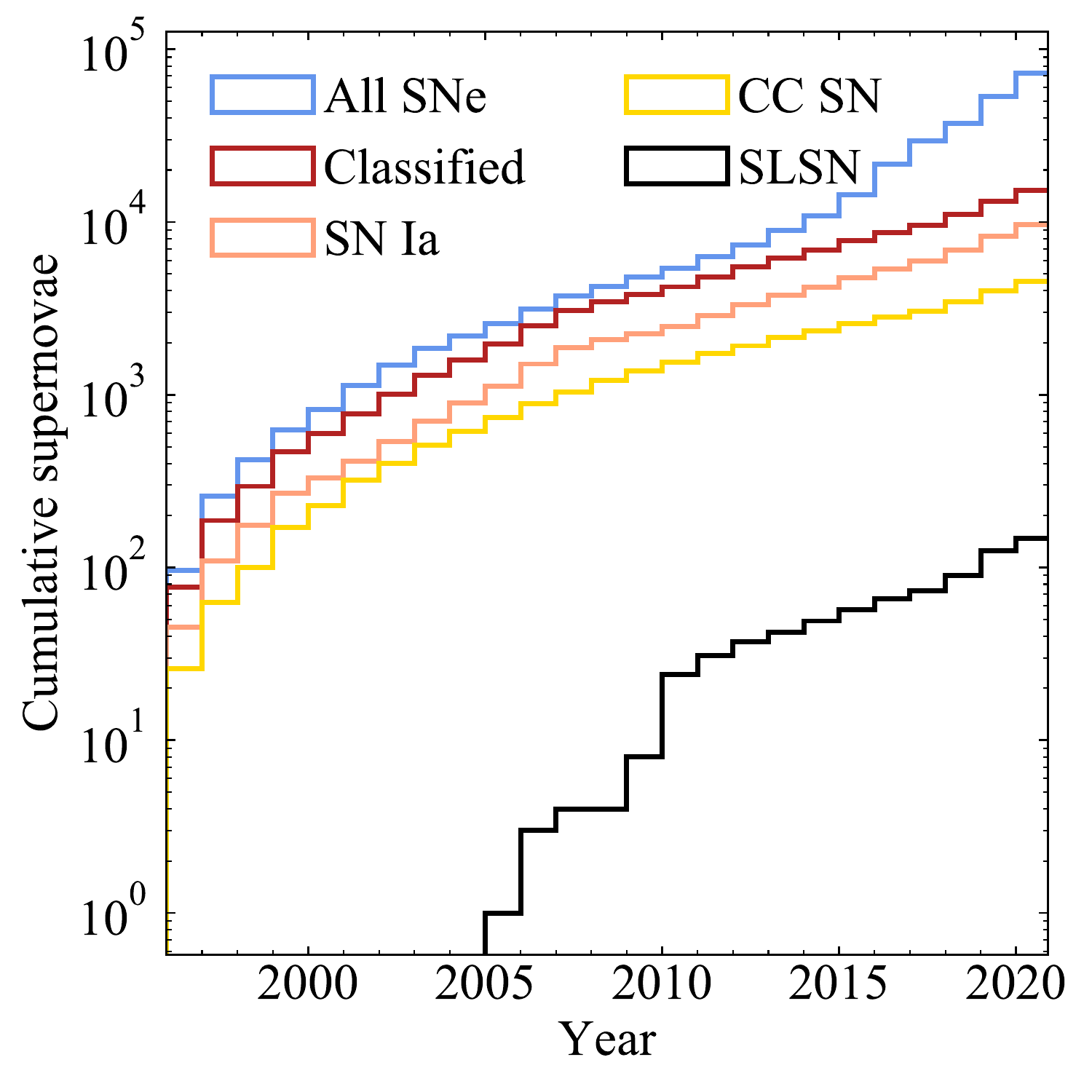}
    \includegraphics[width=5.8cm]{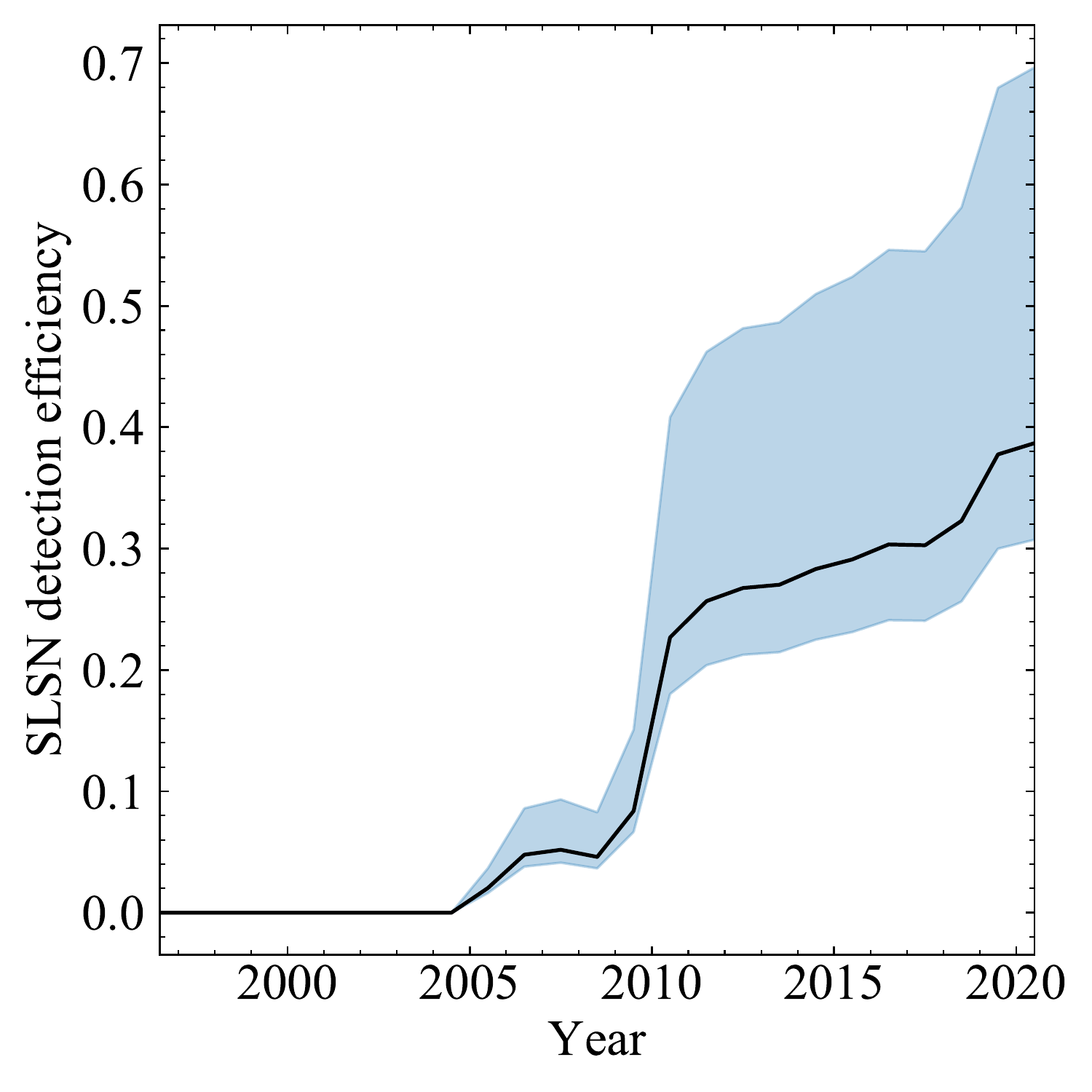}
    \caption{Supernova detection rates and the discovery of SLSNe. Data have been obtained from the \href{https://www.rochesterastronomy.org/snimages/index.html}{Latest Supernova Page}, the \href{https://sne.space/}{Open Supernova Catalog} and the \href{https://www.wis-tns.org/}{Transient Name Server}. Left: Stacked histogram showing supernova discoveries per year, labelled with the survey that detected the majority. The 2010s show a sharp rise driven by robotic surveys. Most events now go unclassified due to finite telescope time. Middle: Cumulative supernova counts. Several thousand core collapse and Type Ia SNe, and over 100 SLSNe, have now been classified. Right: SLSN detection efficiency, defined as the fraction of SLSNe out of all classified supernovae each year, compared to the fraction expected given their estimated occurrence rate (the shaded region shows the uncertainty on their true rate; \citealt{Frohmaier2021}). The efficiency rose dramatically when wide-field surveys removed the bias towards massive galaxies.}
    \label{fig:surveys}
\end{figure*}

A more extreme class of `superluminous supernovae' (SLSNe), $\sim10-100$ times brighter again, escaped our notice until the 21st century. Their eventual discovery was intimately tied with advances in optical sky surveys in the mid-to-late 2000s. Figure \ref{fig:surveys} shows the growth in the supernova discovery rate from 1996-2020, with a sharp uptick around 2010 driven by wide-field robotic telescopes and automated source detection employed by surveys such as the Catalina Real-time Transient Survey \citep[CRTS;][]{Drake2009}, the Panoramic Survey Telescope and Rapid Reponse System \citep[PanSTARRS;][]{Kaiser2010} and  the Palomar Transient Factory \citep[PTF;][]{Rau2009}. Nearly 20,000 supernovae were reported in 2020 -- a 100-fold increase in just 20 years. The volumetric rate of SLSNe is only $\sim$1 in every few thousand supernovae \citep{Quimby2013,Frohmaier2021}, though because these brighter explosions can be detected out to greater distances, they should make up $\sim$1\% of the supernovae detected by a given survey. Finding SLSNe in significant numbers therefore requires the discovery and classification of hundreds or thousands of supernovae per year.

Yet Figure \ref{fig:surveys} also shows that the \emph{fraction} (not just the absolute number) of SLSNe within a given survey also increased dramatically in 2010. Earlier surveys had targeted massive galaxies that have a correspondingly high rate of CCSNe and SNe Ia, but it turns out that SLSNe exhibit a strong and surprising predilection for dwarf galaxies (section \ref{sec:hosts}). Therefore the unbiased nature of modern surveys -- observing wide areas rather than specific galaxies -- was also key to their discovery. While the first reported SLSNe -- SN2005ap \citep{Quimby2007}, SN2006gy \citep{Smith2007a,Ofek2007}, SN2007bi \citep{Gal-Yam2009,Young2010} -- were all major discoveries, the birth of SLSNe as a full-fledged field of study can arguably be traced to 2011. \citet{Quimby2011} presented five SLSNe from PTF, and showed that they formed a class with similar properties to SN2005ap (and the previously mysterious high-redshift transient SCP06F6 \citealt{Barbary2009}). In the same year, \citet{Chomiuk2011} published the first two SLSNe from PanSTARRS. Since then, our knowledge of SLSNe and our ability to spot new candidates have grown hand-in-hand, and we now have a sample of $\sim 100$ such events -- growing rapidly thanks to current surveys like the Zwicky Transient Facility \citep{Bellm2019,Lunnan2020}, ATLAS \citep{Tonry2018}, Gaia \citep{Hodgkin2021} and ASASSN \citep{Shappee2014}.

Here I review the progress we have made in characterising and understanding SLSNe over the past decade, and look ahead to a bright future. The interested reader can find further reviews by \citet{Howell2017}, \citet{Moriya2018}, \citet{Inserra2019}, \citet{Gal-Yam2019a} and \citet{Chen2021}.

\section{Observed and defining properties}

\begin{figure*}
    \centering
    \includegraphics[width=8.8cm]{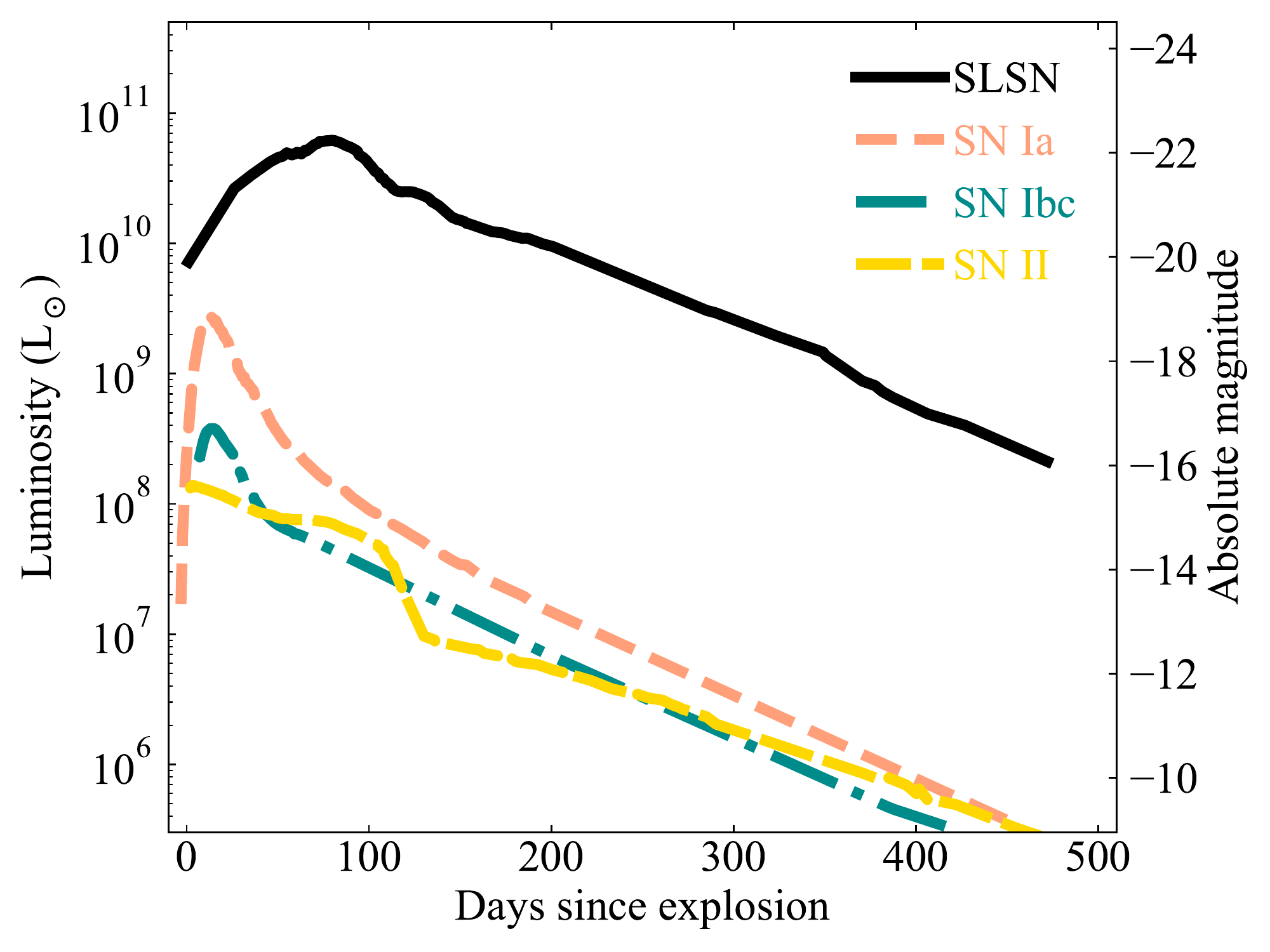}
    \includegraphics[width=6.6cm]{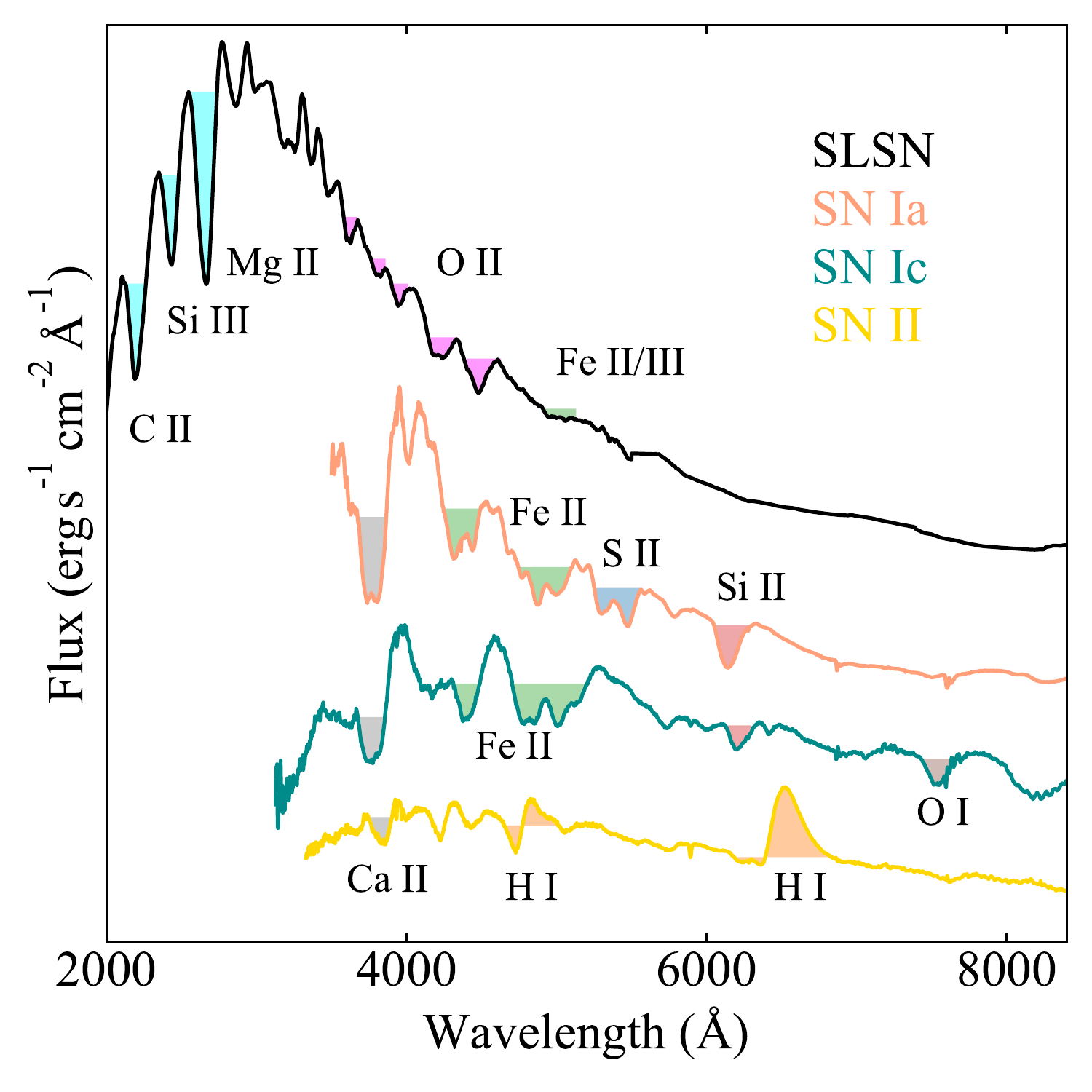}
    \caption{Basic properties of SLSNe compared to SNe Ia and hydrogen-rich (SNe II) and -poor (SNe Ib/c)  CCSNe. Left: The broad, luminous light curves of SLSNe release $\sim 1000$ times more energy than conventional CCSNe. Right: SLSNe are defined by a blue spectrum at the time of peak luminosity, with absorption lines of singly-ionized oxygen. SLSNe show stronger lines in the UV, but still radiate significantly more UV flux than other supernova types.}
    \label{fig:cartoon}
\end{figure*}

\citet{Gal-Yam2012} initially separated SLSNe from other supernova types with a simple threshold in optical absolute magnitude: $M<-21$\,mag, or a luminosity
\begin{equation}
    L_{\rm peak}\gtrsim10^{10}\,{\rm L}_\odot \approx 3\times10^{43}\,{\rm erg}\,{\rm s}^{-1}.
\end{equation}
This is around 10 times brighter than a SN Ia or 100 times brighter than a CCSN (Figure \ref{fig:cartoon}). The time taken to reach peak luminosity is also longer for SLSNe, with exponential rise timescales $\sim 15-50$ days \citep{Nicholl2015}. Integrating the high luminosity over the broad light curve gives a total \emph{radiative} energy $\sim 10^{51}$\,erg, 1000 times that of a normal CCSN and comparable to their total \emph{kinetic} energy.

As with traditional supernovae, SLSNe have been partitioned into two sub-types: with (SLSNe II) and without (SLSNe I) hydrogen lines in the spectrum. Physically, the distinction is whether the star retained a hydrogen envelope until the point of explosion, or was stripped by e.g.~stellar winds or binary interaction. SLSNe II usually show hydrogen emission lines with narrow Doppler widths, indicating low-velocity circumstellar material (CSM) that has been shock-excited by a collision from the supernova ejecta (e.g. \citealt{Smith2007a,Drake2010,Chatzopoulos2011,Benetti2014}; but see \citealt{Miller2009,Gezari2009,Inserra2018} for examples with broad lines). These seem to be differentiated from less luminous Type `IIn' (for narrow-lined) SNe primarily by the radius/density/mass of the CSM, though the extremes of this population require a kinetic energy in the supernova ejecta well in excess of the usual $10^{51}$\,erg \citep{Rest2011,Nicholl2020}.

Hydrogen-poor SLSNe I, on the other hand, have a unique spectrum at the time of their peak luminosity (Figure \ref{fig:cartoon}), with a steep blue continuum and a series of broad absorption lines that \citet{Quimby2011} identified as singly-ionized oxygen (O II). The optical range may also include weak lines of Si II \citep{Inserra2013}, Fe III \citep{Leloudas2012,Nicholl2013} and C II \citep{Nicholl2016,Anderson2018}. The ultraviolet range is more line-rich, showing deep absorptions from Mg II, C II, Si III and Fe III \citep{Quimby2011,Vreeswijk2014,Yan2017a,Yan2018,Quimby2018}. Rather than using a hard magnitude cut, the term SLSN (often used interchangeably with SLSN I) is now typically applied to any event exhibiting this distinctive spectrum, and in fact the range of observed SLSN peak magnitudes now extends at least a magnitude fainter than the original $-21$\,mag \citep{Quimby2018,DeCia2018,Lunnan2018,Angus2019}. There is a continuing debate in the literature as to whether SLSNe I separate further into a class of slowly-evolving events with lower velocities and cooler temperatures, and faster-evolving events with higher velocities and temperatures \citep{Quimby2018,Inserra2018a}. Recently, helium has also been identified in the spectra of up to $\sim10$\% of SLSNe \citep{Quimby2018,Yan2020}, suggesting another possible sub-division into SLSNe Ib (stripped of hydrogen) and Ic (stripped of hydrogen and helium), in analogy with ordinary SNe Ib and Ic.

\begin{figure*}
    \centering
    \includegraphics[width=8cm]{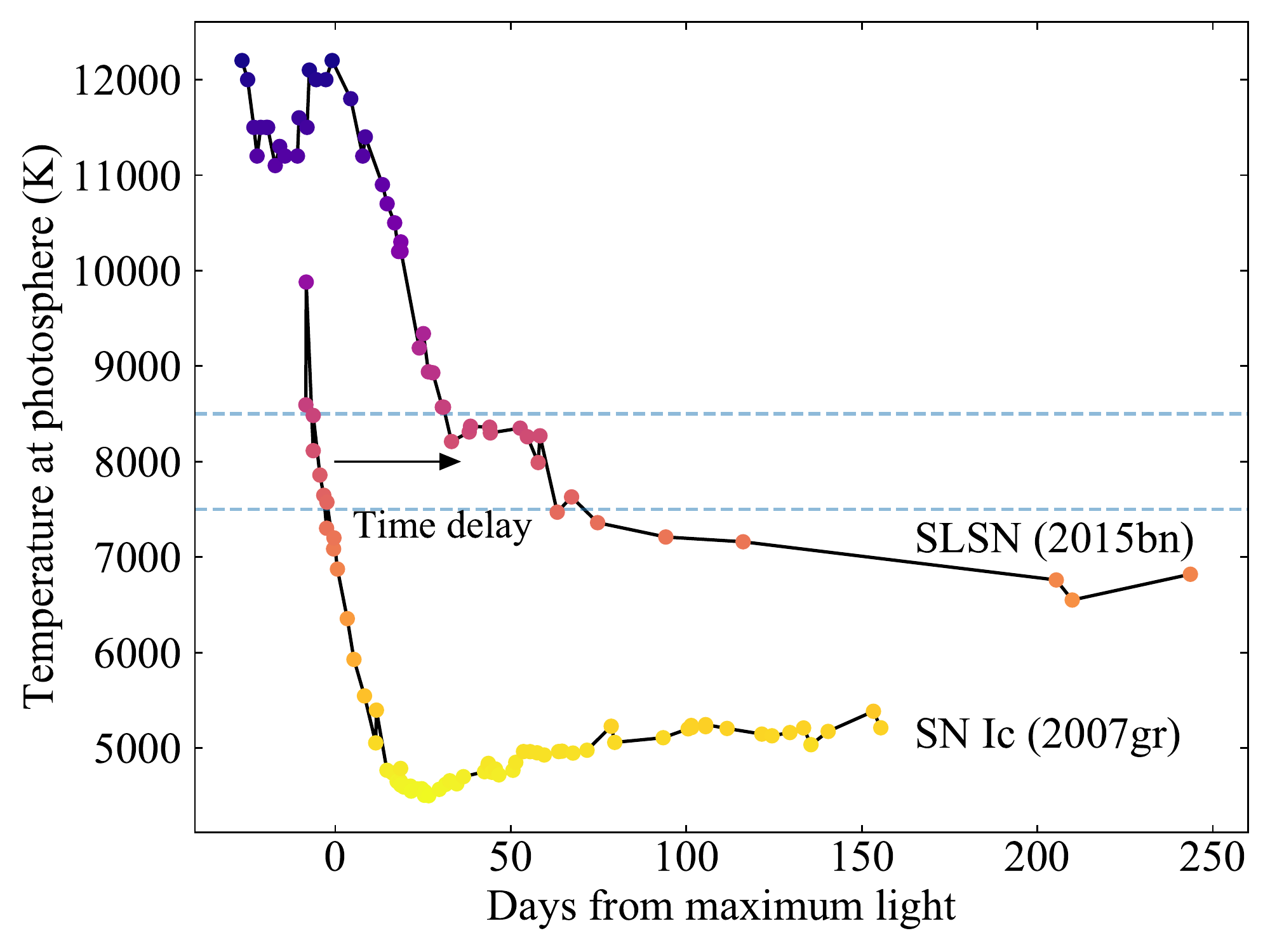}
    \includegraphics[width=8cm]{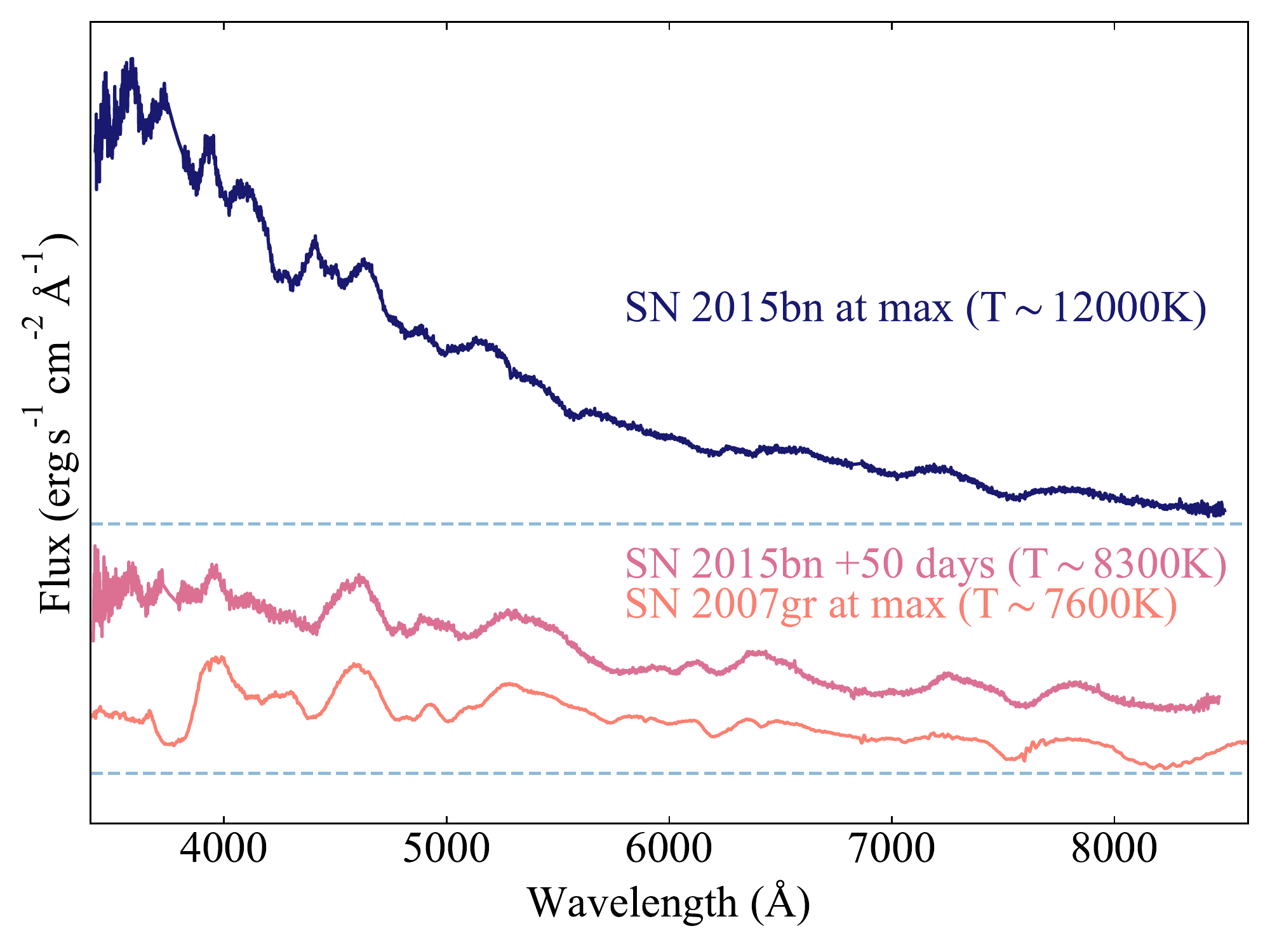}
    \caption{The link between SLSNe and normal stripped-envelope SNe Ic. Left: Temperature evolution. SLSNe are hotter at peak and take longer to cool than SNe Ic. Right: Observed spectra, coloured using the temperature scale from the left panel. Although the spectra of SLSNe and SNe Ic are quite distinct at the time of maximum light, they look remarkably similar when grouped by temperature rather than light curve phase. Data are from \citet{Valenti2008,Nicholl2016a}.}
    \label{fig:temperature}
\end{figure*}

In fact, the relationship between SLSNe and normal stripped-envelope CCSNe is more than analogous. \citet{Pastorello2010} observed that around a month after peak luminosity, the spectrum of a SLSN closely resembles SNe Ic, including the broad-lined SNe Ic that accompany long-duration gamma-ray bursts (long GRBs). This has been confirmed in many studies since, for both fast and slowly evolving SLSNe \citep[e.g.][]{Inserra2013,Nicholl2013,Blanchard2019}. This convergence shows the importance of temperature in determining which atomic lines are excited. SLSNe remain hotter for longer than normal SNe Ic, with $T\gtrsim12000-15000$\,K at maximum light (Figure \ref{fig:temperature}). That SLSNe resemble normal SNe Ic once they have cooled to $\sim 8000$\,K suggests a similar elemental composition in their ejecta: primarily intermediate mass elements such as oxygen, carbon, magnesium and silicon, with some iron-group elements. For this reason I suggest that a better way to think of SLSNe is as `ultra-hot' SNe Ic. The higher temperature at maximum light not only explains the unusual blue spectrum, but accounts for the higher luminosity via Stefan's law of blackbody emission: $L=4\pi R^2 \sigma T^4$, where $R\sim 10^{15}$\,cm is the radius of the emitting surface and $\sigma$ the Stefan-Boltzmann constant. Our goal in explaining the luminosity of SLSNe is thus to uncover this heating source.

\section{Physical mechanisms for SLSNe}

The basic properties of a supernova can be understood using ``Arnett's Rule'' \citep{Arnett1982}. At the peak of the light curve, the luminosity is equal to the heating rate: $L_{\rm peak} = \dot{E}(t=t_{\rm peak})$. The peak occurs roughly on the photon diffusion timescale,
\begin{equation}
    t_{\rm peak} \sim t_{\rm diff} = \left(\frac{3\kappa M_{\rm ej}}{4\pi c v_{\rm ej}}\right)^{1/2},
\end{equation}
where $c$ is the speed of light. The velocity of the ejecta, $v_{\rm ej}\approx 10^4$\,km\,s$^{-1}$, can be measured from Doppler shifts and broadening of spectral lines. The opacity, $\kappa\approx0.1$\,cm$^2$\,g$^{-1}$, is dominated by scattering from free electrons in the ionised ejecta. Finally, $M_{\rm ej}$ is the mass of the supernova ejecta, corresponding to the final mass of the (stripped) star minus the mass of the compact remnant (neutron star or black hole) left behind. Observed SLSN peak times imply a broad range of ejected masses, from a few up to a few tens of solar masses \citep{Inserra2013,Nicholl2015,Nicholl2017,Lunnan2016,Blanchard2020}. 

Three candidate power sources have emerged to explain the long-term heating in SLSNe\footnote{For compact (stripped) stars, the thermal energy deposited by the explosion itself becomes negligible before maximum light, due to adiabatic cooling}, outlined below. Example light curves from these models are shown in Figure \ref{fig:weirdos}.

\subsection{Radioactive decay}

Normal hydrogen-poor supernovae are powered by radioactive $^{56}$Ni, a product of explosive nucleosynthesis, providing an energy 
\begin{equation}
    \dot{E}_{\rm Ni} \sim 10^{43}\ (M_{\rm Ni}/{\rm M}_\odot)\ \left( 6.5 \exp{\left(-t/\tau_{\rm Ni}\right)} + 1.5 \exp{\left(-t/\tau_{\rm Co}\right)} \right)\ {\rm erg}\,{\rm s}^{-1},
\end{equation}
where $M_{\rm Ni}$ is the mass of nickel, decaying with lifetime $\tau_{\rm Ni}=8.8$\,days to $^{56}$Co, which in turn decays with lifetime $\tau_{\rm Co}=111.3$\,days to stable $^{56}$Fe \citep{Nadyozhin1994}. SNe Ic produce $M_{\rm Ni}\lesssim0.1$\,\M\ \citep{Drout2011,Prentice2019}; SNe Ia produce $M_{\rm Ni}\lesssim0.5$\,\M\ \citep[e.g.][]{Childress2015}. To power a SLSN with $L_{\rm peak}\sim10^{44}$\,erg\,s$^{-1}$ at $t_{\rm peak}\sim 30$\,days would need an enormous $M_{\rm Ni}\gtrsim7$\,\M. This could theoretically be produced by a `pair-instability' supernova from a carbon-oxygen core of over 100\,\M\ \citep{Barkat1967,Heger2002}.

However, this 100\,\M\ ejecta would be inconsistent with the observed rise timescales of nearly all SLSNe \citep{Kasen2011,Dessart2013,Kozyreva2015}. Including the requisite nickel mass in a lower mass ejecta would need $M_{\rm Ni}/M_{\rm ej}\sim 1$. Yet the observed spectra of even the most slowly evolving (high mass) SLSNe show primarily carbon-oxygen ejecta, rather than enhanced iron-group absorption \citep{Dessart2012,Nicholl2013}. Finally, observed upper limits on $^{56}$Co luminosity at $\sim 100$ days after explosion are inconsistent with the much larger $^{56}$Ni mass needed to power the peak \citep{Inserra2013,Blanchard2018}. For these reasons, the radioactive model is disfavoured for (at least) the vast majority of SLSNe.

\subsection{Central engine}

A more successful model for explaining SLSNe is a `central engine', with the most obvious candidate being rotation of the newborn neutron star formed during core collapse. Neutron stars have a radius $\sim 12$\,km, giving a minimum spin period $\sim 1$\,ms or a maximum rotational energy $\sim 10^{52}$\,erg (for a typical mass of $1.4$\,\M). If the spin and magnetic axes are mis-aligned, this energy is extracted by magnetic dipole emission  \citep{Ostriker1971,Kasen2010,Woosley2010}:
\begin{equation}\label{eq:mag}
    \dot{E}_{\rm mag} \sim 2\times10^{47}\ (B/10^{14}{\rm G})^2\ (P/{\rm ms})^{-4}\ (1+t/\tau_{\rm mag})^{-2}\ {\rm erg}\,{\rm s}^{-1},
\end{equation}
for initial spin period $P$ in seconds and magnetic field $B$ in Gauss, and a spin-down timescale $\tau_{\rm mag}\approx {\rm 1~day}\ (P/{\rm ms})^2\ (B/10^{14}{\rm G})^{-2}$. Although the energy available from a millisecond rotator exceeds that in a typical supernova, it can only alter the peak luminosity if it spins down in days or weeks, rather than years or longer. The maximum $L_{\rm peak}\sim 10^{45}$\,erg\,s$^{-1}$ occurs for $\tau_{\rm mag}\sim t_{\rm peak}$, which requires $B\sim10^{13-14}$\,G. 

Observed Galactic neutron stars with $B\gtrsim10^{14}$\,G, known as `magnetars', demonstrate that such field strengths are feasible \citep{Thompson1996}. Although known magnetars currently rotate with $P_{\rm now}\gg1$\,ms, their spins may have been much faster at birth. Conservation of angular momentum suggests that many neutron stars should be born rapidly rotating. SLSNe would therefore occur for the fraction of neutron stars that are also born with the right magnetic field.

An alternative central engine utilizes black holes rather than neutron stars. In this scenario, fallback accretion onto the black hole could liberate gravitational energy to power the explosion \citep{Dexter2013}. However, the long durations of SLSNe require fallback timescales more consistent with extended envelopes than with compact carbon-oxygen cores. Furthermore, systematic comparison of this model with observed SLSNe suggests that (for a reasonable conversion efficiency of mass to energy) an unrealistic fallback mass of tens or hundreds of solar masses would be required \citep{Moriya2018a}.

\subsection{Circumstellar interaction}

The final class of models for powering SLSNe is via interaction with a dense CSM, produced by pre-explosion mass-loss from the progenitor. This offers at least three ways to generate a luminous light curve: shock breakout inside an extended wind \citep{Chevalier2011}; diffusion from a shock-heated CSM shell \citep{Smith2007}; and continuous conversion of kinetic to thermal energy as ejecta collide with slower material \citep{Ginzburg2012,Chatzopoulos2012}. In all cases, the relevant energy scale is $\epsilon E_{\rm k}$, with kinetic energy $E_{\rm k} \approx 10^{51}$\,erg and an efficiency factor $\epsilon$, close to unity in SLSNe, for converting this to light. An inelastic collision removes a fraction $\epsilon = M_{\rm CSM}/(M_{\rm ej}+M_{\rm CSM})$ of the ejecta kinetic energy, and hence the condition for powering a SLSN is a CSM mass $M_{\rm CSM}\gtrsim M_{\rm ej}$. This can also be seen via the peak luminosity,
\begin{equation}\label{eq:csm}
    L_{\rm peak,CSM} \sim 1/2\ M_{\rm CSM} v_{\rm ej}^2 / t_{\rm peak};
\end{equation}
SLSNe are possible for $M_{\rm CSM}\gtrsim 1$\,\M. The ejecta must catch up to the CSM before $t_{\rm peak}$, placing it at a maximum distance of $\sim 10^{15}$\,cm. This is consistent with the observed blackbody radii of SLSNe at peak.

The ejecta-CSM interaction model is appealing for a number of reasons. Interaction is seen to varying degrees in many normal supernovae -- exciting narrow lines in the spectra of SNe IIn \citep{Schlegel1990} and producing X-ray and radio emission in both SNe II and SNe Ib/c \citep{Chevalier1994,Weiler2002,Berger2003} -- and probably powers hydrogen-rich Type II SLSNe. 
Moreover, the progenitors of hydrogen-poor SLSNe have lost their outer layers prior to explosion, providing a natural source of CSM. 

However, the radius of the CSM at the time of explosion suggests it was ejected only in the last few years--decades, requiring mass-loss rates $\gtrsim 0.1$\,\M\,yr$^{-1}$. This is orders of magnitude greater than the densest stellar winds from Wolf-Rayet stars \citep{Crowther2007}. The envelope could instead be lost through binary interactions late in the progenitor evolution, or major outbursts: perhaps something akin to the eruptions of luminous blue variable (LBV) stars and/or Eta Carinae, or the predicted `pulsational pair-instability' oscillations of stars with core masses in the range $35-65$\,\M\ \citep{Woosley2007,Woosley2017}.
\\
\\
Given their successes in explaining the luminosities and timescales of 
SLSNe, most studies now focus on central engines and/or circumstellar interaction as the primary energy source(s). However, it remains possible that some other, currently unknown mechanism may turn out to be the key to producing SLSNe. In the next section, I will review how we use other clues to understand the progenitors and explosion mechanisms of SLSNe.

\begin{figure*}
    \centering
    \includegraphics[width=7cm]{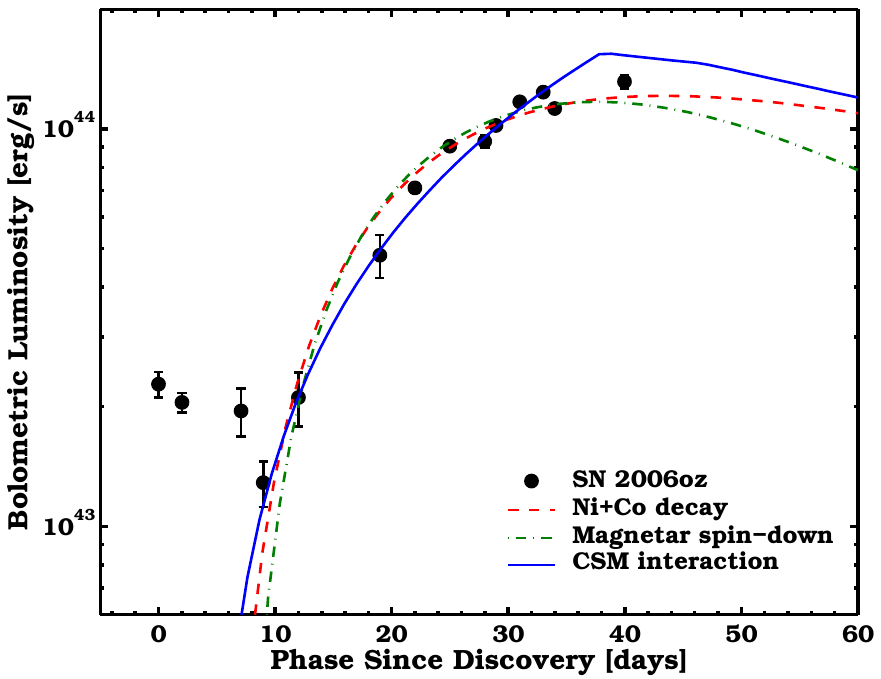}
    \includegraphics[width=9cm]{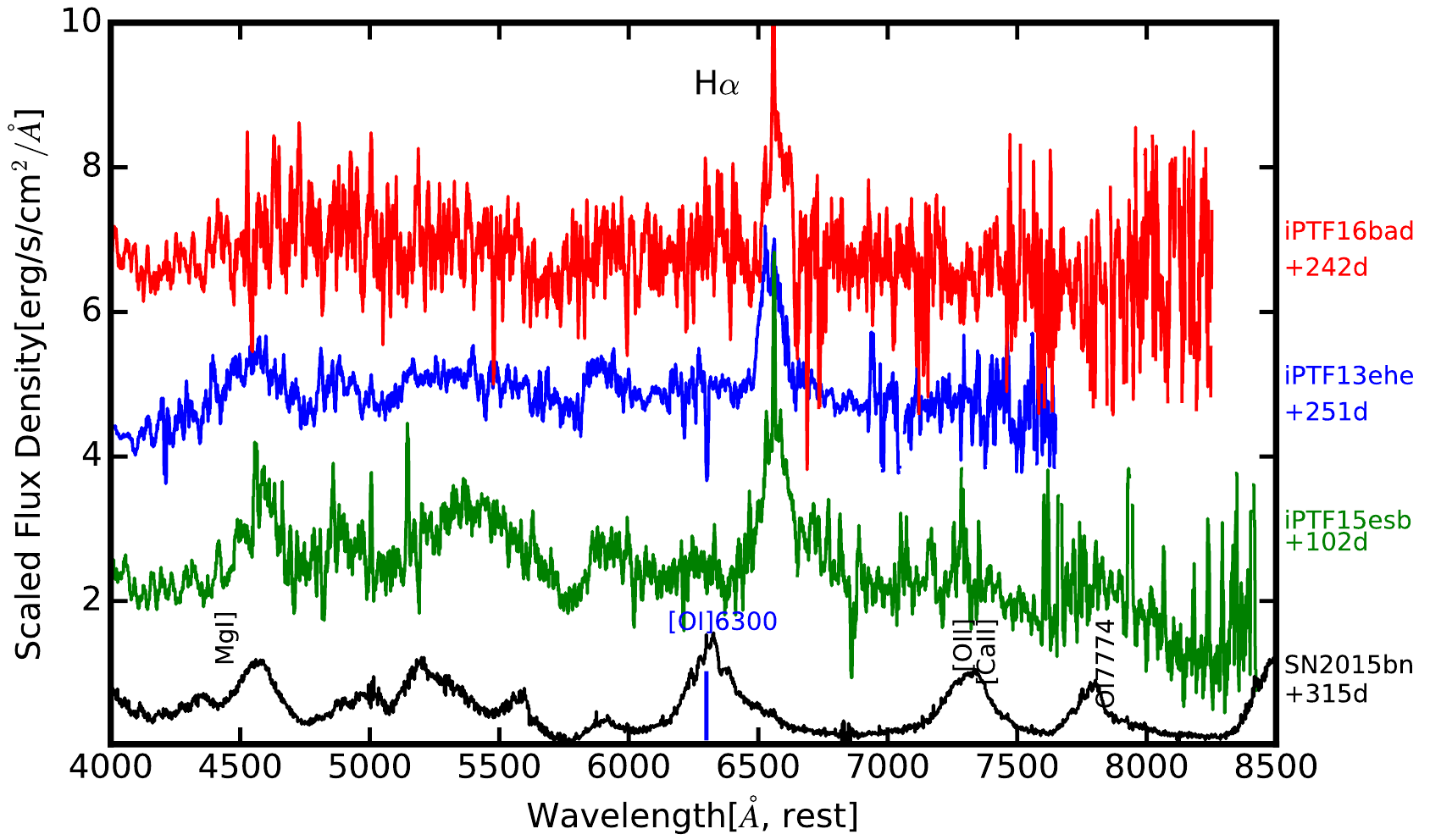}
    \caption{Left, from \citet{Leloudas2012}: Light curve of SLSN 2006oz. Radioactive decay, magnetar spin-down, and circumstellar interaction can all fit the peak luminosity, but the radioactive model requires an unrealistic fraction of $^{56}$Ni in the ejecta. The light curve shows an early `bump', not naturally accommodated by any simple model. Right, from \citet{Yan2017}: Hydrogen Balmer (H$\alpha$) emission lines have appeared suddenly in the spectra of three previously H-poor SLSNe, indicating the ejecta have caught up with a hydrogen envelope lost by the progenitor only decades prior to explosion.}
    \label{fig:weirdos}
\end{figure*}

\section{Observational clues to the nature of SLSNe}

\subsection{Spectroscopy and velocity evolution}


Spectral lines in supernovae are formed by atomic transitions in cooler outer layers, absorbing and scattering blackbody photons emitted from the photosphere (the surface where the optical depth to free electron-scattering is unity). The lines are Doppler-broadened by the expansion of these layers. The similarity with SNe Ic shows that the outer atmospheres must be similar between these classes. Normal SNe Ic are heated by $^{56}$Ni decay, which is produced primarily in the deepest (densest) part of the ejecta. Although we have ruled out this power source in most SLSNe due to the unrealistic nickel fractions required, the spectroscopic similarity likely indicates that the \emph{site} of heating is the same, i.e.~internal to the rapidly expanding line-forming region. This would appear consistent with a central engine, and indeed simulated spectra of centrally-heated carbon-oxygen ejecta have provided good matches to SLSN data \citep{Dessart2012,Howell2013,Mazzali2016,Jerkstrand2017,Dessart2019}.


Circumstellar interaction, in the other hand, is generally considered an external process, where the heating occurs above the fast-moving ejecta as it plows into the CSM. In this picture, the spectral lines are diluted relative to the blackbody continuum, a phenomenon known as top-lighting \citep{Branch2000}. We might also see narrow \emph{emission} lines arising from shock excitation in the CSM \citep{Schlegel1990}. Although the optical absorption lines from SLSNe are shallow compared to those in other supernova types, the strong UV lines show similar equivalent widths across different SLSNe, which is difficult to reconcile with top-lighting by interaction \citep{Nicholl2017c,Yan2018}. Extensive searches are yet to find narrow emission lines in SLSN spectra. However, CSM interaction could remain a viable power source for SLSNe if the CSM has a low covering fraction (for example, an equatorial ring or clumps) and is overrun by the ejecta such that the shock becomes embedded and the slow material obscured from certain viewing angles.


In most supernovae, the observed Doppler widths of the absorption lines decrease over time. This is because their ejecta generally have a `homologous' or `Hubble-like' radial velocity profile, $v(r)\propto r$. As the ejecta expand and become less dense, the optical depth decreases and we see line formation from deeper, and therefore slower, parts of the ejecta. An early testable prediction of magnetar models was that the central pressure would overwhelm the initial density profile and sweep up the ejecta up into a dense shell, keeping the apparent velocity constant in time. With the availability of more advanced modelling in two and three dimensions, the ejecta shell can be smeared out by turbulent mixing, reducing the strength of this effect \citep{Chen2016,Suzuki2019}. However, line velocities in SLSNe do seem to evolve more slowly than in SNe Ic \citep{Nicholl2015,Inserra2018a} (but they do evolve; \citealt{Liu2017,Quimby2018}). Moreover, \citet{Gal-Yam2019} found that O II absorption line profiles in some SLSNe indicated a narrow velocity spread in the ejecta, consistent with a fast, thin shell.


One of the more surprising results in recent years is the discovery that some SLSNe suddenly exhibit hydrogen emission lines at $\sim100-200$ days after explosion (Figure \ref{fig:weirdos}), despite having been previously hydrogen-poor \citep{Yan2015,Yan2017,Chen2018}. This indicates that the ejecta, having expanded to a radius $\sim10^{16}$\,cm, have caught up with the hydrogen envelope lost by the star prior to explosion. The presence of hydrogen-rich material within this radius seems to be limited to $\lesssim20\%$ of SLSNe \citep{Nicholl2019}, but in these events the envelope must have been lost as recently as decades before explosion. Assuming the helium layer was also expelled, it would reside even closer, and its interaction with the ejecta could contribute to the SLSN luminosity at earlier times closer to the light curve peak. In one event, a CSM shell at even larger radius, $\sim10^{17}$\,cm, was detected via an echo of the supernova light \citep{Lunnan2018a}, suggesting envelope loss over a range of time scales.

\subsection{Bumps in the night}


Another surprising finding is the `bumpiness' of SLSN light curves, in contrast to most normal SNe Ic that rise and decline smoothly (although SNe Ia have a well-known secondary maximum in their near-infrared light curves). The first pre-maximum bump for a SLSN was identified by \citet{Leloudas2012}, and is shown as a $\sim 10$ day peak or plateau in Figure \ref{fig:weirdos}. Two more prominent bumps were identified by \citet{Nicholl2015a} and \citet{Smith2016}, prompting \citet{Nicholl2016b} to suggest that such bumps might be a common feature of SLSNe. As predicted, the deep imaging of the Dark Energy Survey proved decisive in testing, and ultimately refuting, this hypothesis, with only three of 14 SLSNe showing clear early-time bumps \citep{Angus2019}.

Several models have been put forward to explain these bumps, including a recombination wave in the ejecta \citep{Leloudas2012} and post-shock cooling of some extended material around the progenitor \citep{Nicholl2015a,Piro2015}. In the context of interaction models, an apparent bump could be caused by a dip, rather than excess, in luminosity, due to an increase in opacity as the CSM becomes ionised \citep{Moriya2012}, however this would require a high covering fraction of CSM and hence may be incompatible with the spectral line velocities. Magnetar-based explanations have also been put forward, including an enhanced magnetar-driven shock breakout \citep{Kasen2016} or a wind driven from the ejecta as an engine-powered jet breaks through \citep{Margalit2018a}.


Complexity in the light curves on longer timescales is even more common, usually in the form of `undulations' during the declining phase (visible in Figure \ref{fig:cartoon}). These seem to be particularly common in longer duration SLSNe \citep{Inserra2017}, but have also been seen in faster events \citep{Blanchard2018,Fiore2021}. Again, attempts have been made to accommodate these in both engine and interaction frameworks. One possibility is a delayed breakout of an ionization front driven by engine \citep{Metzger2014}, which could also be detectable in X-rays for a very nearby SLSN. However, a change in ionization state might be expected to lead to a change in spectral lines, which has not generally been the case observationally. 

Perhaps the simplest explanation for undulations is CSM interaction with an inhomogeneous medium -- either variations in density within a massive CSM that powers the overall light curve, or low-level interaction with a small amount of CSM suddenly encountered by an engine-powered SLSN. In either case, the extra CSM mass needed to supply a luminosity increase of $\sim 10^{43}$\,erg\,s$^{-1}$ on a $\sim10$ day timescale can be estimated from equation \ref{eq:csm} as a few hundredths of a solar mass \citep{Nicholl2016a,Inserra2017}. Such a small mass could perhaps be overrun quickly by the ejecta, without causing significant changes to the spectrum. More dramatic secondary peaks seen in some SLSNe \citep{Vreeswijk2017,Yan2017} may be more challenging to explain in this way.

\subsection{Host galaxies}\label{sec:hosts}

\begin{figure*}
    \centering
    \includegraphics[width=7cm]{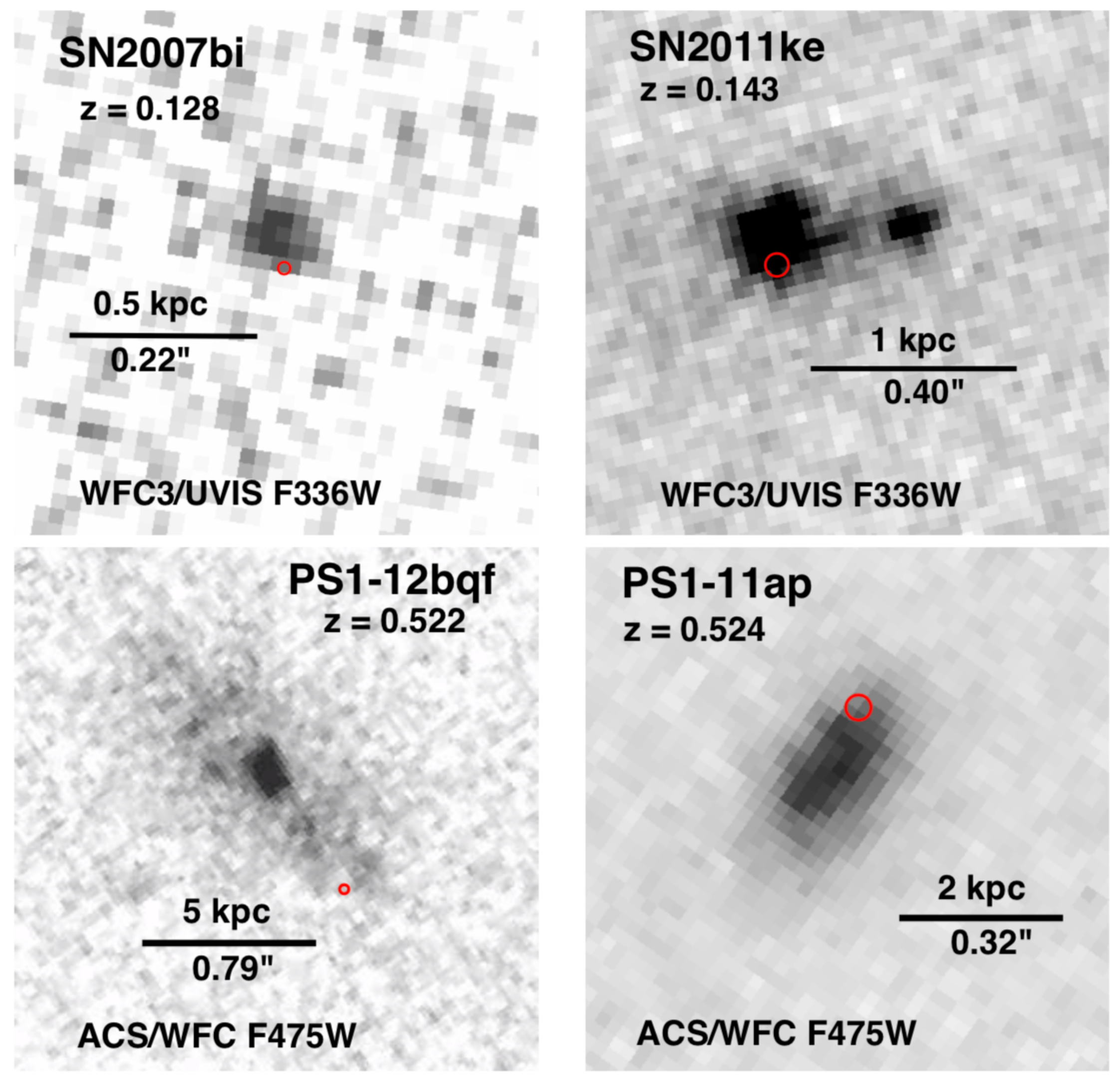}
    \includegraphics[width=9cm]{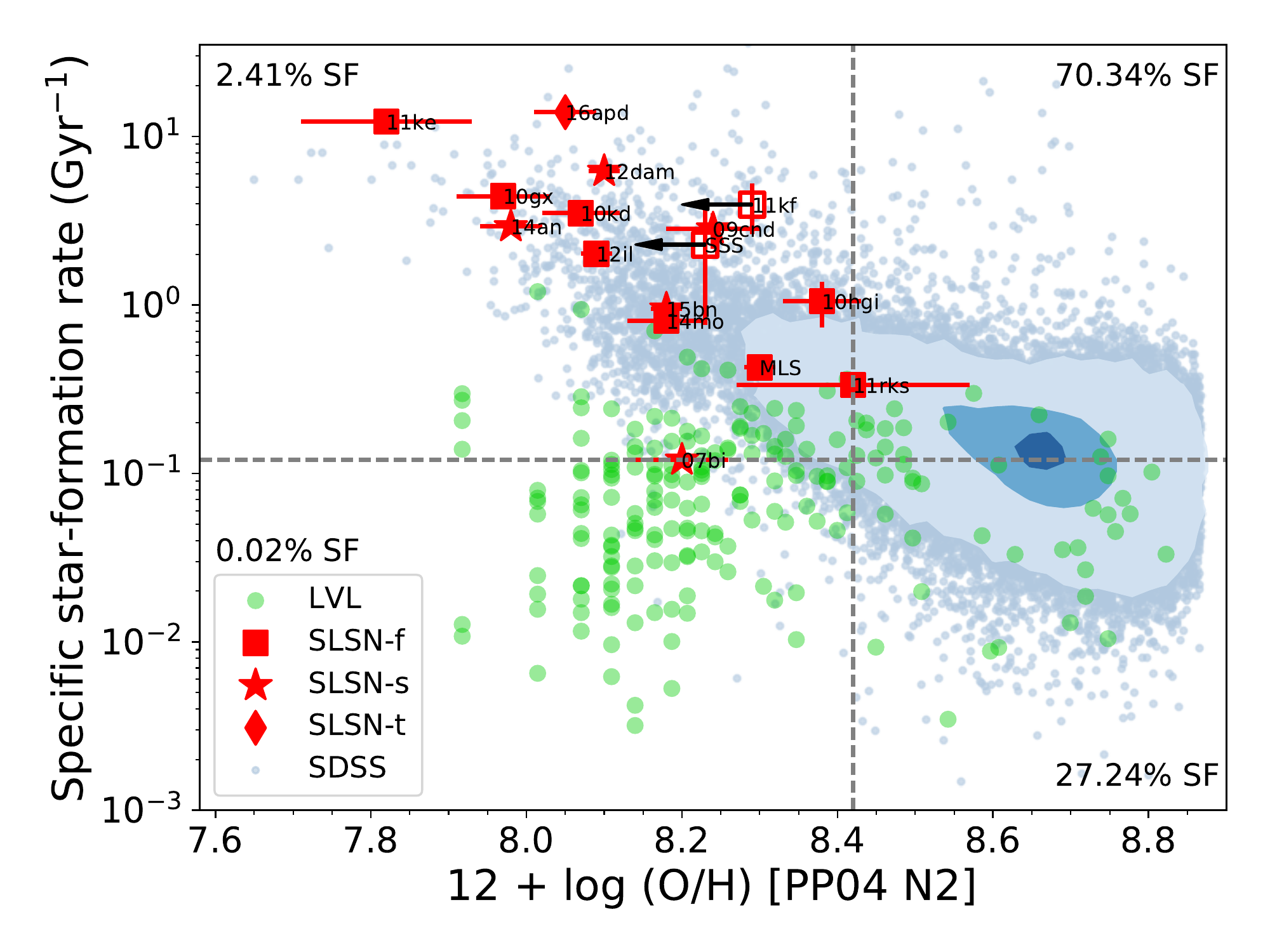}
    \caption{Left, from \citet{Lunnan2015}: \textit{HST} images of four SLSN host galaxies, showing their compact and often irregular morphologies. Right, from \citet{Chen2017a}: Star-formation rate per unit mass versus metallicity (oxygen to hydrogen ratio). Compared to a magnitude-limited sample of galaxies from the Sloan Digitial sky survey, SLSNe occupy galaxies with low metallicity and high specific star-formation rate. In a volume limited sample, only $\approx2\%$ of star-formation occurs in galaxies with metallicities less than $0.5\times$ the solar value, highlighting the importance of this parameter in producing SLSNe.}
    \label{fig:host}
\end{figure*}

In contrast to most supernovae, SLSNe occur almost exclusively in low-mass dwarf galaxies. Example SLSN host galaxies, imaged with \textit{HST} by \citet{Lunnan2015}, are shown in Figure \ref{fig:host}. Since one might naturally have assumed that rare events would be more likely to occur in regions of high stellar mass, this surprising preference for dwarf galaxies must be a clue to their nature. 

Two main physical conditions have been identified in SLSN hosts, and proposed to account for SLSN production: high specific (i.e.~per unit mass) star-formation rate (sSFR), and low metallicities \citep{Neill2011,Chen2013,Lunnan2014,Leloudas2015,Chen2015,Angus2016}. From current samples of SLSN hosts, it appears that the SLSN rate is strongly suppressed at metallicities $\gtrsim 0.5$\,Z$_\odot$, where Z$_\odot$ is the solar ratio of oxygen to hydrogen \citep{Perley2016,Chen2017a,Schulze2018}. As shown in Figure \ref{fig:host}, only 2\% of star-formation in a volume-limited sample occurs at such low metallicity, helping to account for the rarity of SLSNe. However, SLSNe at high metallicity are not impossible, with two (out of $\sim100$ known) having occurred in massive, metal-rich galaxies \citep{Perley2016,Bose2018}. The preference for low metallicities and high sSFR is reminiscent of long GRB host galaxies \citep{Lunnan2014}, although long GRBs appear to have a higher metallicity threshold, $\sim\,$Z$_\odot$, before they are suppressed \citep{Graham2013,Kruler2015,Schulze2018}.

The high sSFRs may be necessary to form very massive stars -- either probabilistically, by sampling the initial mass function more rapidly, or by modifying it to produce a larger fraction of very massive stars. Or, it may simply be an artefact of dwarf galaxies tending to form their stars in bursts \citep{Perley2016}. Many SLSN hosts have been identified as `extreme emission line' galaxies, suggesting a strong ionising radiation field could impact the progenitor evolution \citep{Leloudas2015,Thone2015}. Interactions between SLSN hosts and nearby galaxies may also be important in triggering or modifying star formation \citep{Chen2017b}, with up to 50\% of SLSN hosts having a companion galaxy within 5\,kpc \citep{Orum2020}.

\subsection{Non-thermal emission}

SLSNe emit the bulk of their luminosity thermally, in the UV and optical bands. Searches for non-thermal X-ray and radio emission have provided additional clues to their nature, despite yielding mostly non-detections. \citet{Margutti2018} surveyed 26 SLSNe in X-rays, on timescales from days to years after explosion. They detected the SLSN PTF12dam using the \textit{Chandra X-ray Observatory}, at an X-ray luminosity $L_{\rm X} \approx2\times 10^{40}$\,erg\,s$^{-1}$ around the time of its optical peak, and derived upper limits $L_{\rm X} \lesssim 10^{41}$\,erg\,s$^{-1}$ for virtually all other events. Thus typical SLSNe are at least 100 times fainter in X-rays than in the optical. The only (spectroscopically-normal) SLSN to violate this rule is SCP06F6, with $L_{\rm X} \sim 10^{45}$\,erg\,s$^{-1}$ \citep{Levan2013}. It was suggested by \citet{Metzger2014} that the X-rays in SCP06F6 may have occurred as a result of the engine-powered ionization breakout discussed above.

The story is similar in the radio. \citet{Coppejans2018} compiled all SLSN radio observations at that time (nine events), with no detections. The deepest limits, $\nu L_\nu\lesssim {\rm few}\times 10^{36}$\,erg\,s$^{-1}$, are for the very nearby SN2017egm \citep{Bose2018}. \citet{Eftekhari2021} observed 36 SLSNe with the Very Large Array (VLA) and Atacama Millimeter/Submillimeter Array (ALMA). Their VLA detection of PTF10hgi at 7.5 years after explosion, with $\nu L_\nu\approx 6\times 10^{37}$\,erg\,s$^{-1}$ at 6\,GHz, is the only radio detection in their sample \citep{Eftekhari2019}, and this event remains the only SLSN detected in the radio. The observed emission was consistent with either a pulsar wind nebula energised by a magnetar, or an off-axis jet spreading into our line of sight (in either case, a sign of a central engine). The magnetar hypothesis is supported by additional studies from \citet{Law2019,Mondal2020}. SLSN ejecta is expected to become optically thin to magnetar-powered radio emission only after years to decades, so this picture can be confirmed by continuing to observe the known SLSNe as they age \citep{Margalit2018,Omand2018}. Intriguingly, the radio emission from PTF10hgi also resembles the persistent source associated with the first repeating Fast Radio Burst (FRB), FRB121102 \citep{Chatterjee2017}, which resides in a dwarf galaxy similar to SLSN hosts. This has prompted speculation that at least some FRBs originate from the remnants of SLSNe \citep{Metzger2017,Nicholl2017b}, although subsequent FRB localisations have not shown the same host galaxy preferences. No FRB flares have yet been detected during radio observations of SLSNe.


The numerous non-detections of SLSNe in the X-ray and radio have been used to place tight limits on the density of CSM. In normal CCSNe, the non-thermal emission is synchrotron radiation produced as the blast wave passes through the external medium; it is sensitive to the particle density and the energy of the shock \citep{Chevalier1994,Weiler2002}. The non-thermal emission from a normal SN Ic would be too faint to detect at the distances of most SLSNe, however radio limits for SN2017egm showed that it would be at the faint end even for SNe Ic, which have pre-explosion mass-loss rates $\lesssim 10^{-5}$\,\M\,yr$^{-1}$ \citep[e.g.][]{Berger2002}, assuming Wolf-Rayet star wind velocities $\sim1000$\,km\,s$^{-1}$. Similarly, radio limits for SN2015bn from 1-3 years post-explosion ruled out mass-loss rates in the range $\sim 10^{-5}-10^{-3}$\,\M\,yr$^{-1}$, i.e.~excluding a wind significantly more dense than a normal supernova progenitor \citep{Nicholl2018}. X-ray limits have been used to similarly constrain the wind mass-loss rate $\lesssim10^{-5}-10^{-2}$\,\M\,yr$^{-1}$ \citep{Margutti2018}, or the total shocked CSM mass $\lesssim0.1$\,\M\ \citep{Inserra2017}.


In a GRB afterglow, non-thermal emission is produced when jets, driven by a magnetar \citep{Thompson2004} or accreting black hole engine \citep{MacFadyen1999}, shock the surrounding medium. This produces X-ray and radio emission orders of magnitude brighter than in ordinary supernovae, though the luminosity and the timescale for the afterglow to become visible depend on the orientation of the jet relative to the observer \citep{Meszaros1997}. Detecting jets from SLSNe would be one way to confirm that a central engine operates in these events too, especially in light of their similar hosts to long GRBs, and mutual spectroscopic connection to SNe Ic. \citet{Eftekhari2021} rule out even off-axis jets for 10 SLSNe, and greatly restrict the viable parameter space in jet energy and ambient density for several others, suggesting most SLSNe do not launch successful jets. However, this may be unsurprising even in the magnetar-powered scenario: powering a GRB jet requires that the engine rapidly supply its energy on a timescale of seconds, whereas to continuously heat the ejecta and power a SLSN requires that the engine release its energy on a timescale comparable to $t_{\rm peak}$. The discovery of SN2011kl, a supernova approaching SLSN luminosities, in conjunction with a very rare `ultra-long' GRB lasting for $\sim10,000$\,s \citep{Greiner2015} may represent a transitional event where an engine with an intermediate lifetime manages to power both a jet and a luminous supernova \citep{Metzger2015,Gompertz2017,Kann2019}.

\subsection{Polarimetry}


The polarisation of light emitted from a supernova is a probe of its geometry \citep[e.g.~see review by][]{Wang2008}: an asymmetric distribution of material produces a net polarisation. Asymmetries could arise due to e.g.~bipolar outflows (possibly from failed jets), or a non-uniform distribution of CSM. A handful of SLSNe have now been observed using both imaging- and spectro-polarimetry. In the cases of LSQ14mo, SN2018hti, PS17bek and OGLE16dmu, no intrinsic polarisation could be identified above the interstellar polarisation induced by intervening dust \citep{Leloudas2015b,Lee2019,Cikota2018}. However, high quality polarimetric observations of SN2015bn \citep{Inserra2016,Leloudas2017} and SN2017egm \citep{Bose2018,Saito2020} have shown significant polarisation. In both cases, as the photosphere receded to deeper layers of the ejecta, the polarisation increased, indicating that the inner ejecta was more aspherical than the outer ejecta. In the case of SN2015bn, this transition was quite sudden and coincided with light curve undulations and cooling of the spectrum \citep{Leloudas2017}. The increasing asymmetry has been interpreted as a possible sign of central energy injection along a dominant axis \citep{Inserra2016,Saito2020}.

\subsection{Nebular phase observations}

\begin{figure*}
    \centering
    \includegraphics[width=8cm]{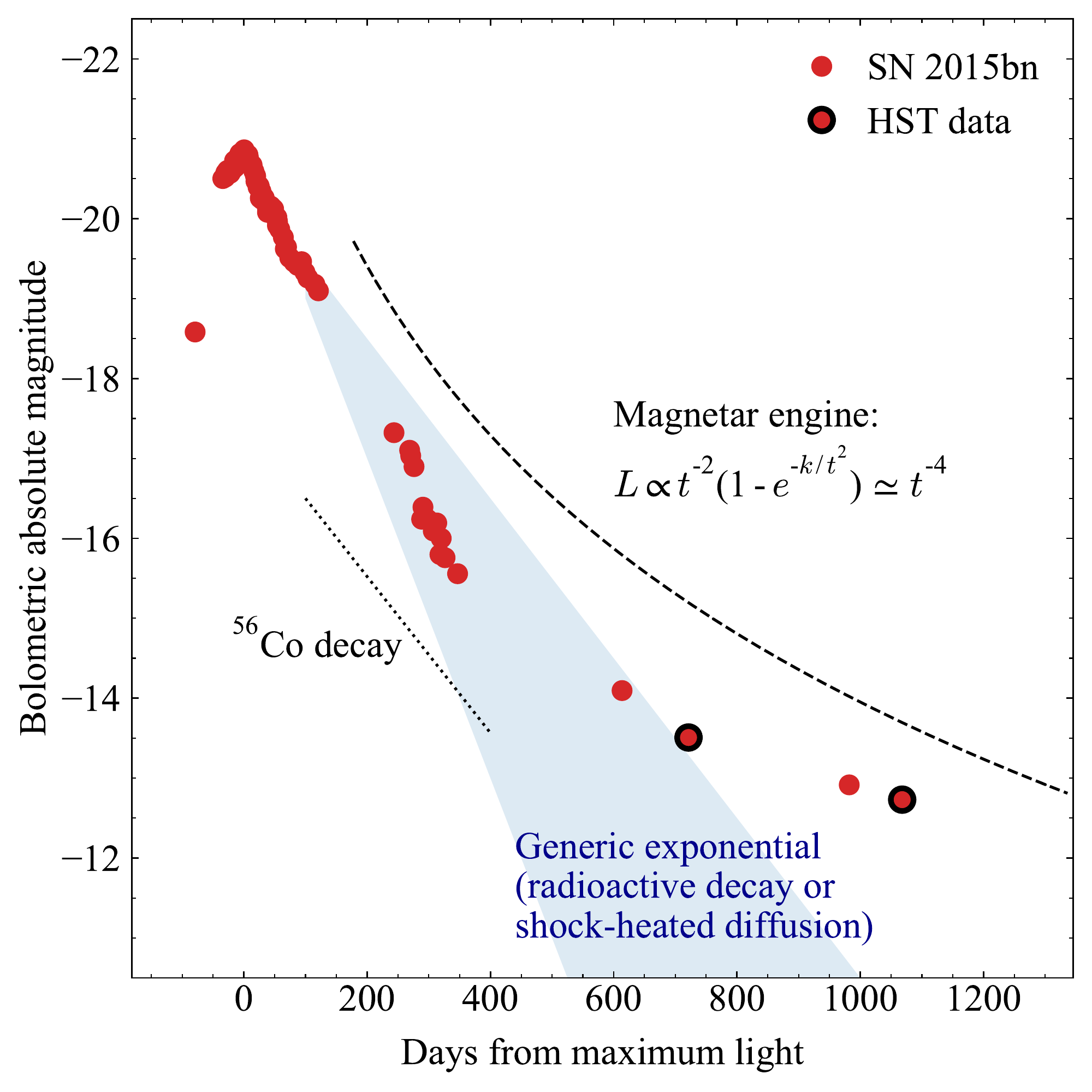}
    \includegraphics[width=8cm]{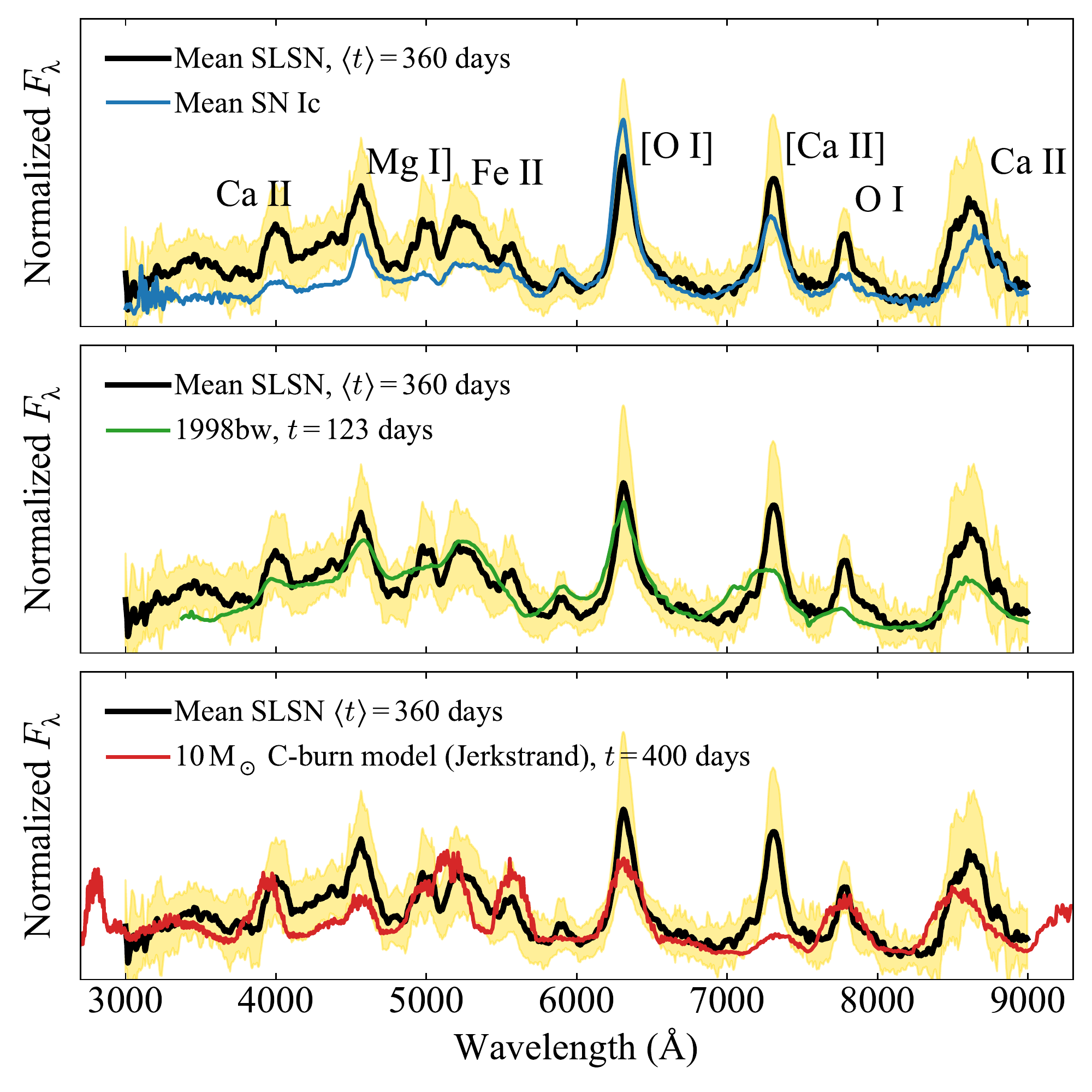}
    \caption{Long-term observations of SLSNe. Left: the light curve slopes predicted by different models diverge at $\sim$ hundreds of days after explosion. Two SLSNe with \textit{HST} imaging in this phase show a flattening consistent with magnetar engines \citep{Nicholl2018,Blanchard2021}. Right: Average spectrum at one year after explosion, when the ejecta have become transparent, compared to normal SNe Ic (top), the GRB-supernova 1998bw (middle) and a model from \citet{Jerkstrand2017} (bottom). The strongest emission lines are labelled; the square-bracket notation indicates `forbidden' lines, i.e. unlikely transitions that can only occur in low-density gas. SLSNe and GRB-supernovae show strong iron lines below $\sim5000$\,\AA, pointing to more massive progenitors than normal SNe Ic. The strong oxygen and calcium lines are consistent with clumpy ejecta and a central heating source \citep{Jerkstrand2017,Nicholl2019}.}
    \label{fig:late}
\end{figure*}


As supernovae expand and cool, the photosphere continues to recede deeper into the ejecta. By a time
\begin{equation}
    t_{\rm neb} \sim 1\,{\rm year}\ (M_{\rm ej}/10\,{\rm M}_\odot)^{1/2}\ (v_{\rm ej}/10^4\,{\rm km}\,{\rm s}^{-1})^{-1},
\end{equation}
the ejecta become fully transparent and the so-called `nebular phase' begins. There is no longer an electron-scattering blackbody continuum or absorption lines. Instead, various (permitted and forbidden) emission lines appear as the electron density drops below critical values \citep[see review by][]{Jerkstrand2017a}. The first nebular spectrum of an SLSN was obtained for SN2007bi by \citet{Gal-Yam2009}. By 2019, the nebular sample had grown to 12 events, with the majority from PTF \citep{Quimby2018}, enabling the first statistical studies \citep{Nicholl2019}. The mean nebular spectrum of SLSNe is shown in Figure \ref{fig:late}. The observed emission lines are the same as in SNe Ic, however the strength of the iron lines from 4000-5000\,\AA\ is significantly enhanced in SLSNe, similar to nebular observations of broad-lined SNe Ic associated with long GRBs \citep{Nicholl2016,Jerkstrand2017}. The O\,I $\lambda$7774 recombination line is stronger in SLSNe than in normal or broad-lined SNe Ic. Its luminosity combined with its narrow velocity width indicates ongoing ionisation in the inner regions of the ejecta, likely requiring continuous central energy injection \citep{Milisavljevic2013,Nicholl2016,Nicholl2019}. 

Model spectra of slowly-evolving SLSNe computed by \citet{Jerkstrand2017} showed that most of the lines can be reproduced by $\sim 10$\,\M of explosive carbon-burning products (which are mostly oxygen by mass). They also determined from calcium line ratios that SLSN ejecta must be significantly `clumped', i.e. consisting of compact dense regions in an otherwise rarified volume. Evidence for clumping was further observed in a larger sample of events, and the electron density within the clumps ($n_e>10^8$\,cm$^{-3}$) helps to account for the strength of the O\,I recombination line \citep{Nicholl2019}. Clumping can arise due to engine-driven pressure instabilities \citep{Kasen2010,Chen2016}, but may also occur due to fragmentation of a cold dense shell at an eject-CSM interface (though \citealt{Jerkstrand2017} note that it could be difficult to encompass enough mass in this shell). An outstanding mystery in SLSNe is why some events show apparently nebular emission from [Ca\,II] much earlier, during the photospheric phase \citep{Gal-Yam2009,Inserra2017} -- this requires some part of the ejecta to become rarified much more rapidly than in conventional supernovae.


Following SLSNe deeper into the nebular phase provides another handle on the heating source. As shown in Figure \ref{fig:late}, radioactivity, magnetar spin-down, and CSM interaction diverge in their predictions for the light curve slope, though distinguishing between them robustly requires observations years after explosion \citep{Woosley2010,Inserra2013,Moriya2017}. Such observations have now been obtained with the \textit{Hubble Space Telescope} (\textit{HST}) for SN2015bn \citep{Nicholl2018} and SN2016inl \citep{Blanchard2021} at phases of $\approx500-1000$ days. The data appear to most closely track the magnetar spin-down scenario, but prefer steeper power-law indices $L\propto t^{-\alpha}$ with $\alpha=3-4$ rather than the fiducial $\alpha=2$ (equation \ref{eq:mag}). This can be explained if a fraction of the engine power `leaks' out of the ejecta without heating it; as the ejecta become less dense, a larger fraction escapes \citep{Wang2015,Chen2015,Nicholl2018}. Deep upper limits in X-rays \citep{Bhirombhakdi2018} and radio \citep{Nicholl2018} at the same phase as the optical detections of SN2015bn showed that $\lesssim1\%$ of the available engine energy could be escaping in these wavelengths. Given the hard spectrum expected from a magnetar, GeV-TeV gamma-rays are the most likely waveband for leakage; upper limits from the \textit{Fermi} satellite have not yet been deep enough to rule this out \citep{Renault-Tinacci2018}. However, recent simulations by \citet{Vurm2021} showed that gamma-ray escape on timescales of $\sim1$ year require low magnetization of the expanding nebula, potentially posing difficulties for the `leaky magnetar' picture.


\section{Diversity and progenitors}

SLSNe appear to originate from stars more massive than typical CCSN progenitors: this is evidenced by their longer durations (and implied ejecta masses), nebular spectra with strong [O\,I] and Fe II emission, and the intensely star-forming environments in which they occur. Based on light curve fits assuming a magnetar model, \citet{Blanchard2020} estimated pre-explosion progenitor masses (corresponding to the carbon-oxygen core of the stripped star) in the range 3.6--40\,\M. A steep drop-off at the top of this range may indicate an upper limit where the progenitor encounters the pulsational pair-instability. In contrast, the progenitors of normal CCSNe mostly fall below $\lesssim 20$\,\M\ \citep[e.g.][]{Smartt2009}. To eject the highest masses inferred in SLSNe, especially at the observed high velocities, likely requires an explosion energy $\sim 10^{52}$\,erg, an order of magnitude greater than the canonical energy scale for normal supernovae \citep{Blanchard2020,Mazzali2016}.

\begin{figure*}
    \centering
    \includegraphics[width=8.3cm]{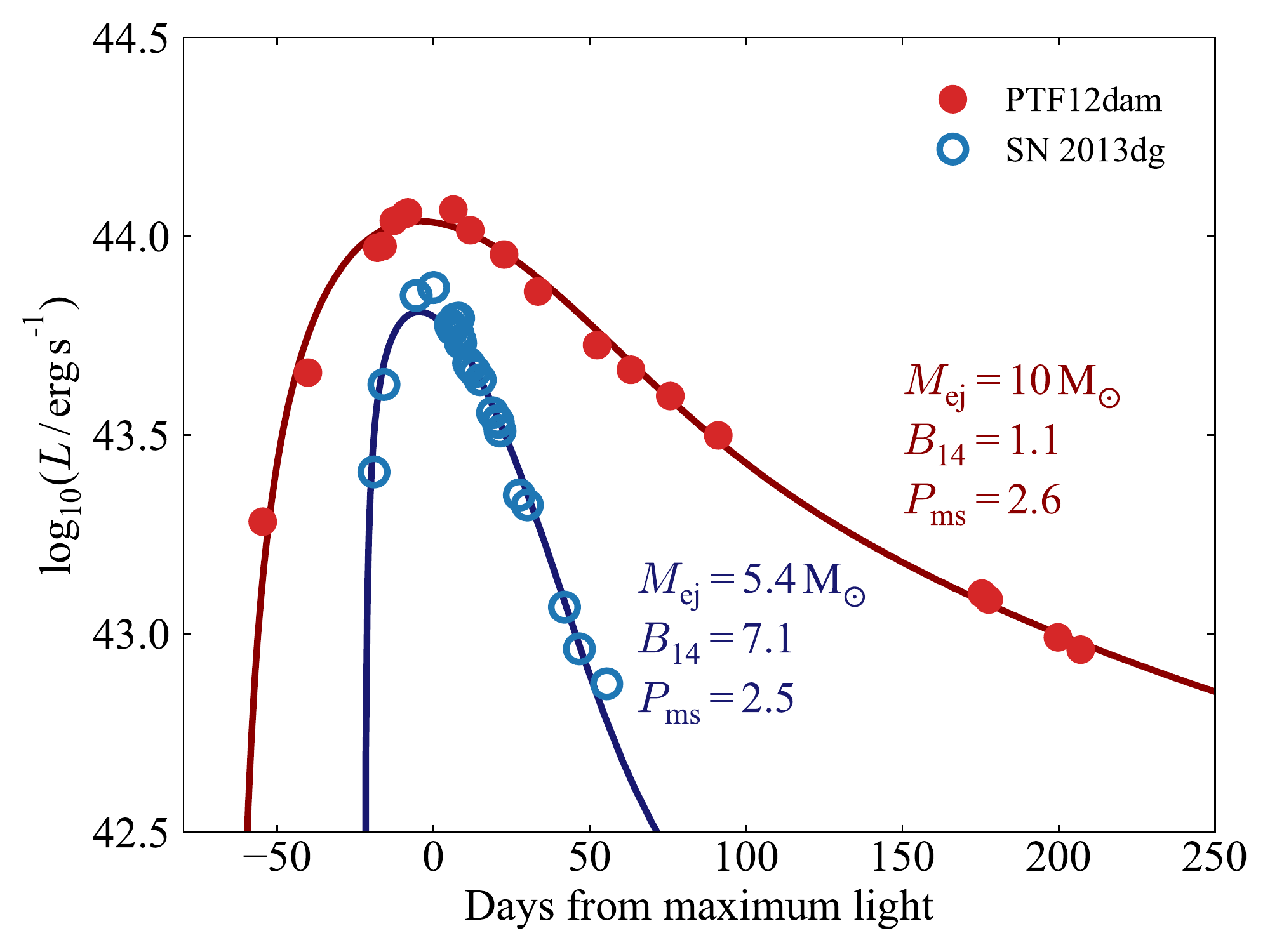}
    \includegraphics[width=7.8cm]{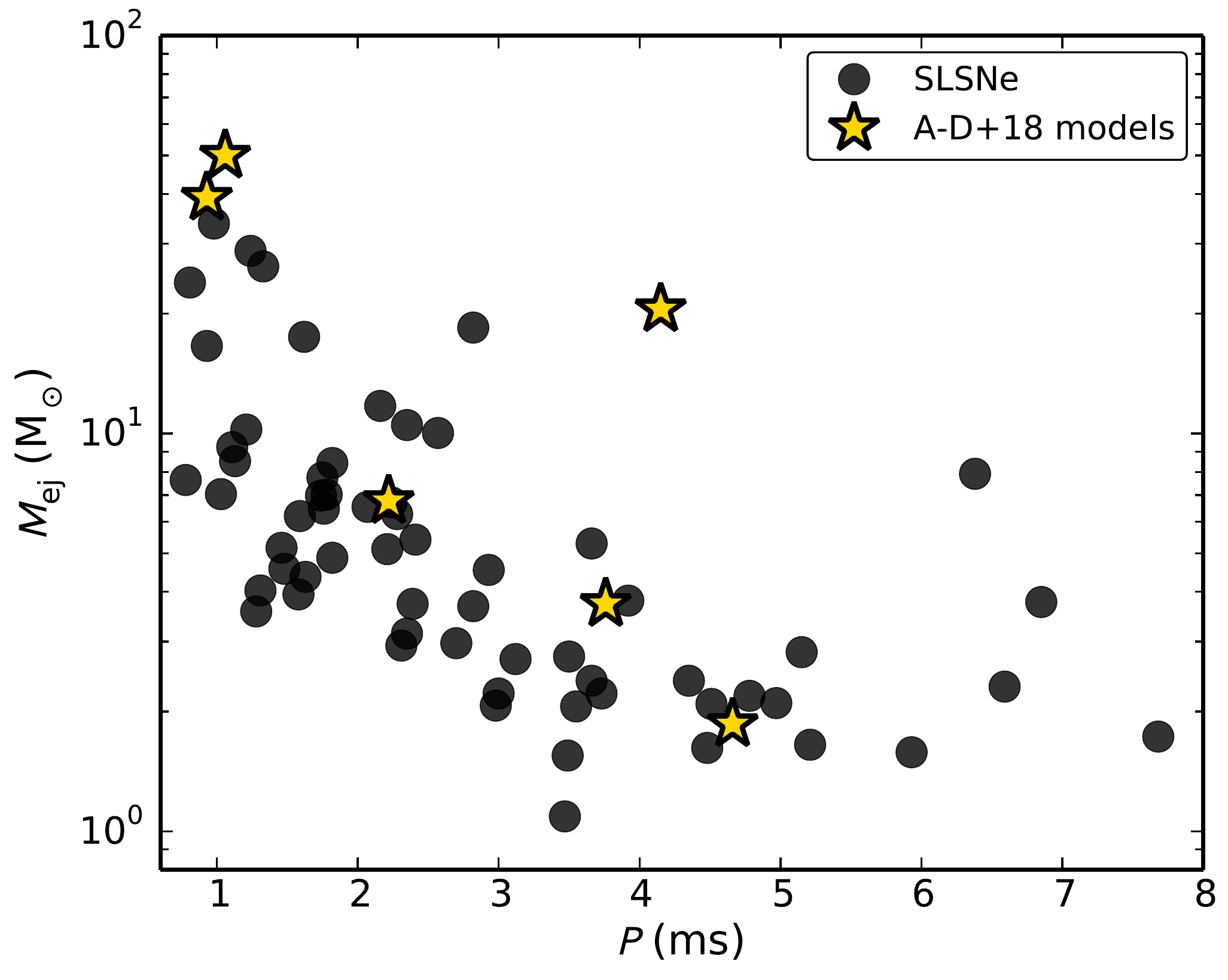}
    \caption{Left: Diverse SLSN light curves fitted with magnetar models. More massive ejecta and lower magnetic fields lead to slower evolution, and the longer rise time results in a larger maximum light radius and a cooler spectrum for the same luminosity. Right, from \citet{Blanchard2020}: Ejecta mass versus magnetar spin period from model fits to 62 SLSNe. The observed correlation could be explained by mass-stripping: progenitors that lose more mass before explosion also lose more angular momentum. The results are consistent with simulations by \citet{Aguilera-Dena2018}.}
    \label{fig:progenitors}
\end{figure*}

The spectroscopic (especially in the nebular phase) and host galaxy connections between SLSNe and long GRBs could indicate that SLSNe have similar progenitors and/or explosion mechanisms to the engine-powered GRBs.
Assuming a magnetar engine, Figure \ref{fig:progenitors} shows how slower evolving SLSNe can be explained by higher ejecta masses and weaker magnetic fields. \citet{Nicholl2017} noted that at least some of the diversity in SLSN spectra at maximum light may be a result of longer rise times in more massive events, leading to a larger photosphere and cooler temperature for a similar luminosity and velocity.
Although the magnetar model has been criticised as overly flexible, in fact it has the minimum number of extra parameters to decouple the peak luminosity from the timescale, and produce more diverse light curves than radioactive models ($P$ and $B$ rather than just $M_{\rm Ni}$), and fewer parameters than CSM interaction models. Given the \emph{observed} diversity in SLSNe, an extra degree of freedom is not arbitrary, but rather a necessary and sufficient physical condition.

\citet{Nicholl2017} suggested that the relative rate of SLSNe compared to normal CCSNe is consistent with the small fractional volume of mass-spin-magnetic field space that produces over-luminous events.
For $B \gtrsim 10^{15}$\,G, the energy would be injected in seconds rather than days, driving a GRB jet rather than a SLSN \citep{Metzger2015}. For $B\lesssim 10^{13}$\,G, the magnetar spins down too slowly to boost the energy or peak luminosity of the supernova, producing only a normal SN Ic. In this case evidence of the magnetar could still appear in the late-time light curve, as was possibly seen in the SN Ic iPTF15dtg \citep{Taddia2019}. A continuum of magnetic field strengths (or spin periods) predicts a continuum of luminosities between the various classes of stripped supernovae, and in fact this is now observed in unbiased supernova samples \citep{DeCia2018}.

If central engines are the key ingredient for making SLSNe, their progenitors must be rapidly rotating. Simulations show that if the core is spinning fast enough at collapse to make a millisecond neutron star, a dynamo mechanism can naturally amplify the magnetic field to the range required for SLSNe \citep{Duncan1992,Mosta2015}. The low-metallicity environments where most events are located help to reduce angular momentum loss from metallicity-dependent, line-driven stellar winds \citep[e.g.][]{Vink2021}. However, since SLSNe (and GRB-supernovae) lack hydrogen, they must have lost their envelopes somehow. \citet{Woosley2017} found that rapidly-spinning single-star models with initial masses in the range 60--100\,\M\ can retain high core angular momentum even after losing their envelopes via pair-instability pulsations. Rotation also leads to more massive cores by mixing in fresh material from the envelope. Interaction with a close binary companion could offer an alternative pathway to remove the envelope while maintaining (or even spinning up) the rotation of the progenitor \citep[e.g.][]{Yoon2010,deMink2013}. \citet{Stevance2021} identified plausible binary progenitor models for the SLSN 2017gci, but could find no viable single star scenarios within their library of population synthesis models.

The connections between binary interaction, rotation, core mass and envelope loss may underpin much of the diversity in SLSNe. \citet{Chen2017a} identified a possible correlation between magnetar spin periods (from model fits to SLSN data) and host metallicity, which could be explained by more mass-stripping in metal-rich environments, though this was not recovered in other studies \citep{Nicholl2017,DeCia2018}. \citet{Blanchard2020} found an anti-correlation between ejected mass and spin period, consistent with more heavily stripped stars having lost more angular momentum (Figure \ref{fig:progenitors}). This correlation has also been found by \citet{Aguilera-Dena2018,Aguilera-Dena2020} in their simulations of rapidly rotating massive stars. These models also predict $\sim 0.5$\,\M\ of CSM close to the progenitor, due to rotational and neutrino-driven mass-loss during late burning stages. This mass of CSM, and the timing of ejection, is compatible with the undulations in some SLSN light curves. A consistent picture may be emerging in which differing degrees of CSM interaction add another layer of complexity to SLSN observables, on top of the underlying variation from different ejecta and engine parameters.

\section{Looking ahead}

\begin{figure*}
    \centering
    \includegraphics[width=10cm]{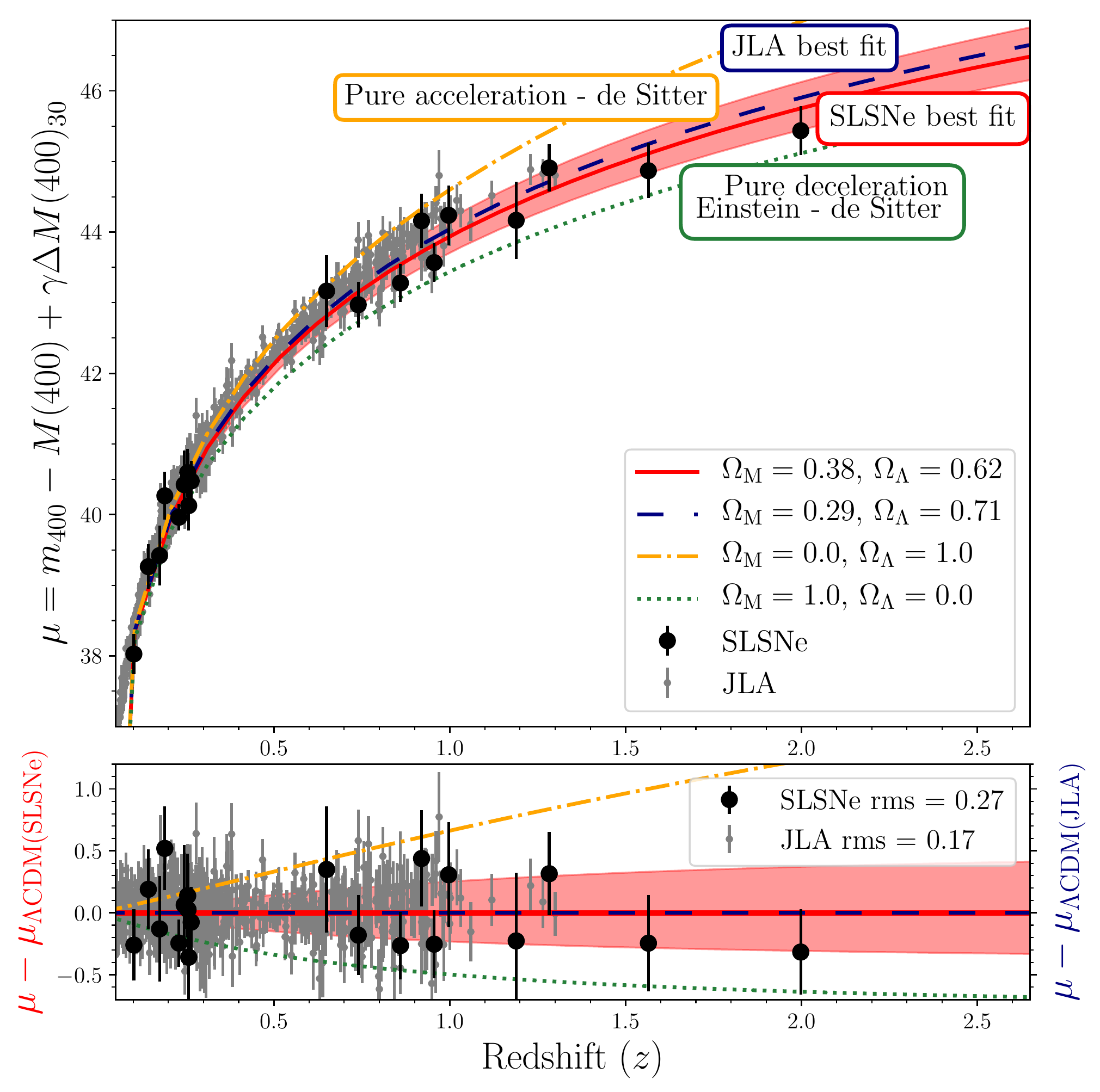}
    \caption{From \citet{Inserra2021}: Hubble diagram (distance vs redshift) for SLSNe (black) and SNe Ia (grey). Supernova cosmology relies on determining the true peak luminosity of an event from a distance-independent parameter such as light curve width (a `standardisable candle'). Comparing the observed and intrinsic brightness gives the distance. The redshift, derived from the spectrum, measures how quickly the Universe is expanding at that distance. Because of their luminosity, SLSNe can be detected at much greater distances than SNe Ia. If they can be standardised, they enable measurements of the expansion history of the Universe at earlier times when differences between cosmological models are more pronounced.}
    \label{fig:cosmo}
\end{figure*}

We are now entering the statistical age of SLSNe. The new Vera Rubin Observatory, conducting the Legacy Survey of Space and Time (LSST), is predicted to find $\sim 10^4$ SLSNe per year \citep{Villar2018}. Machine learning techniques will be required to separate SLSNe from other transients in this enormous data stream \citep[e.g.][]{Gomez2020,Villar2020,Hosseinzadeh2020,Muthukrishna2019}, and enable spectroscopic and multi-wavelength follow-up. With thousands of SLSNe, we will much more finely sample their diversity and energetics, and continue to pin down their progenitors. As the first wide-field survey with an 8\,m telescope, LSST will be especially important in building up the sample at higher redshift. SLSNe are detectable at much greater distances than other transient types, with a handful of known events at redshifts $z\sim2-4$ \citep{Cooke2012,Pan2017,Smith2018,Moriya2019,Curtin2019}. Increasing these numbers will enable us to probe conditions and star formation in the early Universe \citep{Berger2012} and to extend supernova cosmology beyond the reach of SNe Ia \citep[][Figure \ref{fig:cosmo}]{Inserra2014,Inserra2021}.

When it comes to SLSN physics, progress requires a better understanding of how the engine -- if indeed one is present -- releases its energy to the ejecta. This could be in the form of a particle wind, Poynting flux, or magnetic reconnection \citep[e.g.][]{Margalit2018a,Vurm2021}. Detecting or ruling out high-energy leakage at late times will provide important constraints. It is equally important to determine the fraction of SLSNe that interact with CSM, and to establish a more robust mapping between the CSM properties and their effect on SLSN observables, allowing us to trace out the final decades in the lives of these stars via the material expelled during this time. Observations within the first days after explosion, made more feasible by the deep imaging from LSST, will probe material even closer to the progenitor, reveal the nature of the light curve `bumps', and -- hopefully -- produce even more surprises.

\section*{Acknowledgements}

MN thanks Peter Blanchard, Ting-Wan Chen, Cosimo Inserra, Giorgos Leloudas, Ragnhild Lunnan and Lin Yan for use of their figures.
MN is supported by a Royal Astronomical Society Research Fellowship and by the European Research Council (ERC) under the European Union’s Horizon 2020 research and innovation programme (grant agreement No.~948381).





\begin{multicols}{2}
\bibliographystyle{mnras}
\bibliography{refs} 
\end{multicols}


\bsp	
\label{lastpage}
\end{document}